\documentclass[universe,article,accept,pdftex,moreauthors]{Definitions/mdpi}

\newcommand\arcsec{\mbox{$^{\prime\prime}$}}

\firstpage{1} 
\pubvolume{1}
\issuenum{1}
\articlenumber{0}
\pubyear{2024}
\copyrightyear{2024}
\datereceived{ } 
\daterevised{ } 
\dateaccepted{ } 
\datepublished{ } 
\hreflink{https://doi.org/} 

\Title{Observations of the formation of a proto-spot in a pre-existing field environment}

\TitleCitation{Observations of the formation of a proto-spot in a pre-existing field environment}

\Author{Mariarita Murabito $^{1,2,}$*\orcidA{}, Ilaria Ermolli $^{1}$\orcidB{}, Salvo~L. Guglielmino $^{3}$\orcidC{}, Paolo Romano $^{3}$\orcidD{}, Fabrizio Giorgi $^{1}$\orcidE{}}

\AuthorNames{Mariarita Murabito, Ilaria Ermolli, Salvo~L. Guglielmino, Paolo Romano, Fabrizio Giorgi}

\AuthorCitation{Murabito, M.; Ermolli, I.; Guglielmino S.~L.; Romano, P.; Giorgi, F.}

\address{%
$^{1}$ \quad INAF -- Osservatorio Astronomico di Roma, via Frascati 33, I-00078, Monteporzio Catone, Italy\\
$^{2}$ \quad Space Science Data Center (SSDC), Agenzia Spaziale Italiana, via del Politecnico, s.n.c., I-00133, Roma, Italy\\
$^{3}$ \quad INAF -- Osservatorio Astrofisico di Catania, Via S. Sofia 78, I-95123 Catania, Italy}

\corres{Correspondence: mariarita.murabito@inaf.it}

\abstract{Bipolar emerging flux regions (EFRs) form active regions (ARs) that generally evolve in a pre-existing magnetic environment in the solar atmosphere. Reconfiguration of the small- and large-scale magnetic connectivities is invoked to explain a plethora of energy release phenomena observed at the sites of EFRs. These include brightening events, surges, and jets, whose trigger and relationship are still unclear. In this context, we study the formation of a proto-spot in AR NOAA~11462 by analyzing spectropolarimetric and spectroscopic measurements taken by the Interferometric Bidimensional Spectrometer along the Fe~I 630.2~nm and Ca~II 854.2~nm lines on April 17, 2012. We complement these high-resolution data with simultaneous SDO satellite observations.   
The proto-spot forms from magnetic flux emerged into the photosphere that coalesces following plasma flows in its surrounding. The chromospheric and higher atmosphere observations show that flux emergence occurs in a pre-existing magnetic environment, with small- and large-scale coronal arcades that seemingly shape the proto-spot formation in the upper atmospheric layers. In addition, in the chromosphere we observe an arch filament system and repeated intense brightening events and surges, likely due to magnetic interactions of the new flux with the pre-existing overlying coronal field. These phenomena are observed since early stages of the new flux emergence.
}


\begin{document}


\section{Introduction}

The emergence of magnetic flux bundles by magnetic buoyancy is generally believed to be the driving process behind  active regions (ARs) appearance on the solar atmosphere. Indeed, observations of the simplest ARs can be explained in terms of distinct bipolar emerging flux regions (EFRs), while in general as the outcome of several bipoles emerging in close succession \cite{Schmieder:14,vandriel2015,cheung2017}.

In the photosphere, ARs  manifest themselves in the form of dark (sunspot and pores) and bright (faculae) features produced by more concentrated and diluted magnetic fields, respectively, while in the chromosphere and outer atmosphere, ARs appear overall brighter with increasing height.

The formation of ARs is preceded by a clear sequence of events. EFRs appear at the solar surface as small-scale flux elements of kG field strength that move apart. Same-polarity concentrations merge forming pores, which further combine yielding proto-spot and later spots at the outer edges of the EFRs. In the chromosphere, EFRs show up as bright plages and dark filamentary structures seen in the Ca II K and H${\alpha}$ lines, respectively \cite{Yadav:23}. The dark structures, called arch filament systems (AFSs, \cite{Bruzek:67,Bruzek:69}), connect the opposite polarity concentrations observed in the photosphere, crossing the polarity inversion line of EFRs. They are typically observed during the emergence phase \cite{Strous:99,Spadaro:04,Zuccarello:05}, also in penumbra formation sites \cite{Murabito:17}, although they can persist until the decay of magnetic flux concentrations \cite{Contarino:09,Sergio:17}. Finally, in the transition region and corona, bright, dense loops appear above the chromospheric AFSs. 

Generally ARs develop in pre-existing magnetic environments. Therefore,
\begin{quote}
	``as new magnetic flux emerges through the solar atmosphere in the form of $\Omega$--loops, from beneath the solar surface, through the solar photosphere into the chromosphere, corona and beyond, it perturbs local conditions causing  the reconfiguration of the small- and large-scale magnetic connectivities'' \cite{vandriel2015}.
\end{quote}

Such a magnetic rearrangement, which involves reconnection and conversion of the magnetic energy stored in field lines into heat and kinetic energy, is invoked to explain a plethora of phenomena observed above EFRs. These include impulsive energy release processes manifested as localized and intense brightening events, as well as plasma ejections such as surges and jets. At larger scale, it can eventually trigger the ejection of high-velocity, high-temperature clouds of magnetized solar plasma traveling throughout the heliosphere and potentially affecting technological infrastructures and biological systems in it \cite{Schrijver_etal2015}. 

Therefore, ARs evolve depending on the evolution of the assembling EFRs, on their interaction with the surrounding magnetic fields, and on new flux emergence in their vicinity. Moreover, since the early stages of new flux emergence, EFRs start reconnecting with surrounding pre-existing fields at a slow and steady rate, forming new coronal magnetic connections \cite{Tarr:14}. 

In recent years, the dynamics of newly emerged magnetic flux and of its interaction with pre-existing magnetic fields have been investigated with multi-wavelength high-resolution observations, \cite{Guglielmino:10,Santiago:12,Santiago:14,Ortiz:14,Ortiz:16,Shelton:15,Jaime:15,centeno2017,guglielmino2018,verma2018,guglielmino2019}. In the lower atmosphere, these observations display occurrence above EFRs of Ellerman bombs \cite[][]{Georgoulis:02,Pariat:04} and UV bursts \cite{Peter:14,Tian:16,Young:18}, while in the higher atmosphere of X-ray jets \cite{Innes:16,Raouafi:16}, brightening due to quiescent reconnection \cite{Tarr:14}, and outflows at the edges of the EFRs \cite{Harra:10,Harra:12}. Multi-dimensional magnetohydrodynamic (MHD) simulations of magnetic flux emergence have also shed light on the origin and connection of these different phenomena \cite{mactaggart2015,ni2015,ni2016,nobrega2016,nobrega2017,hansteen2017,ni2018,nobrega2018,nobrega2022}. 

However, all of the above  observational studies focus on the atmospheric response to small-scale EFRs, while the simulations on the origin of certain observed phenomena are carried out under simplified approaches, applied in relatively small numerical domains. 

In a previous paper \cite[][hereafter Paper~I]{Ermolli:17}, we analysed the formation of a pore in the photosphere, observed with ground-based high resolution full Stokes measurements and co-temporal space-borne observations since the early stages of an EFRs to the creation of a funnel-shaped structure. The studied pore is a proto-spot of the AR NOAA~11462, since it later developed into the leading spot of that activity centre. The AR, which was characterized by a simple division between opposite magnetic polarities, did not develop flares during its passage on the solar disc, but its evolution seemingly triggered the development of flare active regions in its vicinity.  

The observations analysed in Paper~I show translational and rotational motions of the plasma surrounding the evolving feature, which are reminiscent of the rise of a twisted flux tube. These motions seem to foster the wrapping of the new emerged flux to form a funnel-shaped pore. Co-temporal EUV observations suggest that the pore formed in a pre-existing magnetic environment consisting of a smaller-scale, almost horizontal coronal arcade and of a larger-scale coronal arcade protruding into South. Therefore, we wondered if the pre-existing arcades could affect the evolution of the new emerged flux, and which was their role in the phenomena associated to the evolution of the EFR.

In this paper, we study the interaction of the forming structure with the pre-existing chromospheric and coronal fields  located above and in the vicinity of the evolving region. In particular, we analyse the response of the overlying atmospheric layers to the emerging flux and examine whether and how the large scale geometry of the pre-existing coronal arcade may have step in the pore formation and in the dynamic events observed from the chromosphere to the corona above the EFR. We also discuss the role of the shearing and rotational motions of opposite polarity patches observed during the pore formation as a trigger for the events observed higher in the atmosphere. 

The data set is described in Sect.~2, the plasma motions and magnetic field evolution derived from the photospheric and chromospheric data, as well as brightening events and dynamics obtained from the TR and coronal observations are presented in Sect.~3. Section~4 discusses our results, which are summarized in Section~5.

\section{Data and Methods}

We analysed high-resolution data taken with the Interferometric Bidimensional Spectrometer \cite[IBIS,][]{Cavallini_2006} at the Dunn Solar Telescope of the National Solar Observatory (NSO/DST).

The IBIS observations were taken on April 17, 2012, from 13:58 to 20:43~UT, by targeting the formation of a pore along the central meridian, during the initial evolution of AR NOAA 11462. Note that no IBIS data were acquired between 16:30 UT and 18:30 UT because of a worsening of the seeing. The observations comprise of 223 sequences, each containing  imaging spectropolarimetry along the photospheric Fe~I 617.3~nm and Fe~I 630.2~nm lines, and imaging spectra along the chromospheric Ca~II 854.2~nm line, which is the focus of the present work. Data were taken at 24, 30, and 25 spectral line positions, respectively, with a cadence of 67~s, over a field-of-view (FOV) of $\approx 40 \times 90 \,\mathrm{arcsec}^2$, with a pixel scale of  $\approx 0.09$\arcsec. We limit our analysis to a smaller portion of $\approx 50 \times 30 \,\mathrm{arcsec}^2$, 
 showing the evolution of the pore at the center of the IBIS FOV.  
Further information on the IBIS data, and on the methods applied to perform the standard reduction and the restoring for seeing-induced degradations, is available in Paper~I. 
Co-alignement of photospheric and chromospheric data was applied with respect to the simultaneous white-light image, using cross-correlation techniques during the standard data reduction.
The observations analysed in this study are available at the IBIS data archive \cite[IBIS$-$A,][]{Ermolli2022}. 

High-resolution observations were complemented by data acquired by the Helioseismic and Magnetic Imager \cite[HMI,][]{Scherrer_etal2012} and by the Atmospheric Imaging Assembly \cite[AIA,][]{Lemen_etal2012} aboard the Solar Dynamics Observatory \cite[SDO,][]{Pesnell_etal2012} satellite. 
We considered SDO/HMI and SDO/AIA observations taken on April 17, 2012, from 0:00 to 24:00~UT, some of which simultaneous to the IBIS data. We analysed sub-arrays from the SDO/HMI and SDO/AIA full-disk observations extracted from individual images, as well as the sub-arrays obtained by using standard SDO/HMI Space-weather Active Region Patches \cite[SHARP,][]{Hoeksema_etal2014,Bobra_etal2014}, employing SolarSoft IDL routines. SDO/HMI data consist of photospheric continuum filtergrams and vector magnetograms obtained 
at the Fe~I 617.3~nm, with a pixel size of $\approx 0.5$\arcsec and cadence of 45~s. SDO/AIA images is comprised of filtergrams taken at various passbands: 160~nm, 30.4~nm, 21.1~nm, 17.1~nm, 19.3~nm, 13.1~nm, 33.5~nm, and 9.4~nm lines, with a pixel size of $0.6$\arcsec. The cadence for the UV 160~nm passband is 24~s, while for the remaining EUV passbands is 12~s. These passbands sample the solar atmosphere from the temperature minimum to the corona.

To study the response of the upper atmospheric layers to transient photospheric magnetic activity and plasma flows caused by the proto-spot formation, we created co-aligned SDO and IBIS datacubes. This allowed us to locate the different features and processes observed by the various instruments at different atmospheric heights. We applied cross-correlation techniques, by using the Fe~I 617.3~nm line-continuum observation from both instruments as a reference and the magnetic features as fiducial elements, after re-sampling of IBIS data to the pixel scale of SDO/HMI observations. As a result, we obtained co-aligned time series of IBIS measurements and of SDO/HMI and SDO/AIA images. The subFOV selected for the analysis encompasses the flux emergence site.

We computed the plasma line-of-sight (LOS) velocities in the chromosphere from the shift of the Ca~II line core. The zero-velocity reference was estimated by the core positions of the time-and-space averaged line profiles in the quiet Sun part of the FOV. Chromospheric LOS velocities were averaged on the same $3 \times 3 \,\mathrm{pixel}^2$ box.

In addition, we studied the horizontal plasma motions in the IBIS FOV, estimating their velocity $\mathrm{v}_{H}$ with  the Fourier Local Correlation Tracking method \cite[FLCT,][]{Fisher_welsch2008}. We applied this method to the available line-continuum data for both the photospheric Fe~I 630.2~nm line and the chromospheric Ca~II line measurements. The FWHM of the Gaussian tracking window was set to $0.5$\arcsec, to properly track magnetic structures with spatial scales smaller than the typical granular size. The  temporal integration was made over a 13-minute time interval.

\section{Results}

\subsection{General overview}
\label{GO}
\begin{figure}[H]
	\centering{
		\includegraphics[width=12.5cm, trim=30 80 30 0, clip]{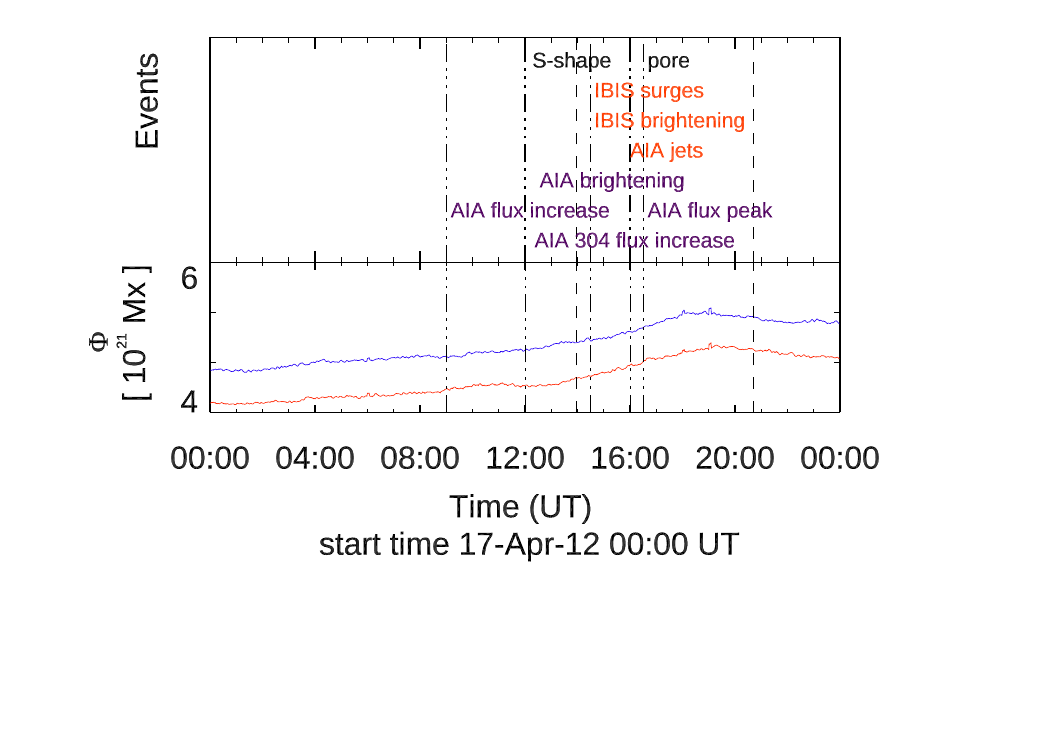}}
		\caption{Evolution of the magnetic flux (bottom panel) in the studied region, derived from SDO/HMI LOS magnetograms acquired on April 17, 2012, from 00:00 to 24:00~UT, by analyzing the entire FOV shown in Figure~\ref{f2}. The red and blue lines indicate the positive and negative magnetic flux, respectively. The dashed vertical lines in all panels indicate the time interval of the IBIS observations. Additional information  indicates occurrence of events described in Section 3 (top panel). Labels refer to events observed in IBIS, SDO/AIA, and SDO/HMI continuum data. The dotted vertical lines mark time the events listed in the top panel are observed for the first time in the analysed observations.\label{f1}}
\end{figure}

Figure~\ref{f1} (bottom panel) displays the evolution of the magnetic flux of the analysed region as deduced from SDO/HMI magnetograms. The time interval of the IBIS measurements and a number of events observed during the pore formation, over a broad range of heights of the solar atmosphere, are marked in the same figure (top panel). Specifically, transient intensity enhancements and surges in the chromospheric Ca~II 854.2~nm data and brightening and jets in the EUV coronal lines, presented in the next sections, are indicated. 

A magnetic flux increase is clearly seen in the region during IBIS observations: it starts a few hours earlier, at around 6:00~UT, and lasts until around 17:00~UT, with small flux changes before and after the period of IBIS observations. 

\begin{figure}[H]
	\begin{adjustwidth}{-\extralength}{0cm}
		\centering
		\includegraphics[width=18.5cm, trim=0 225 130 0, clip]{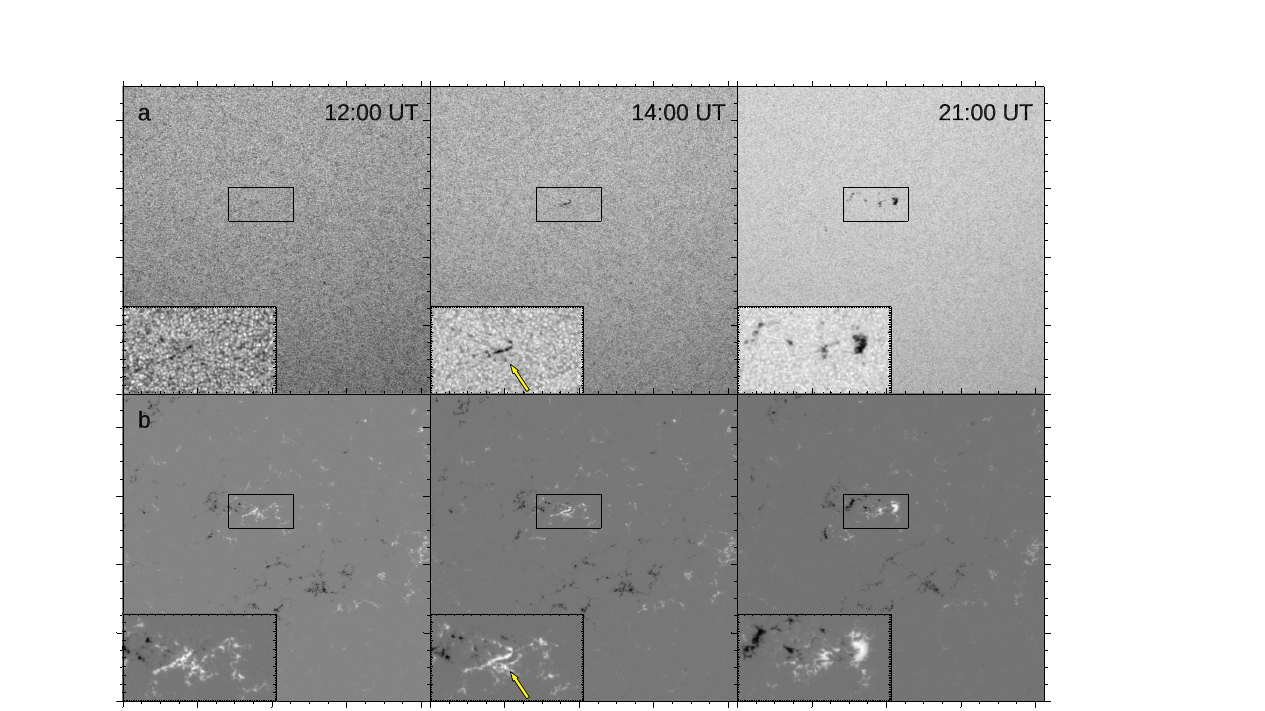}
		\includegraphics[width=18.5cm, trim=0 180 130 40, clip]{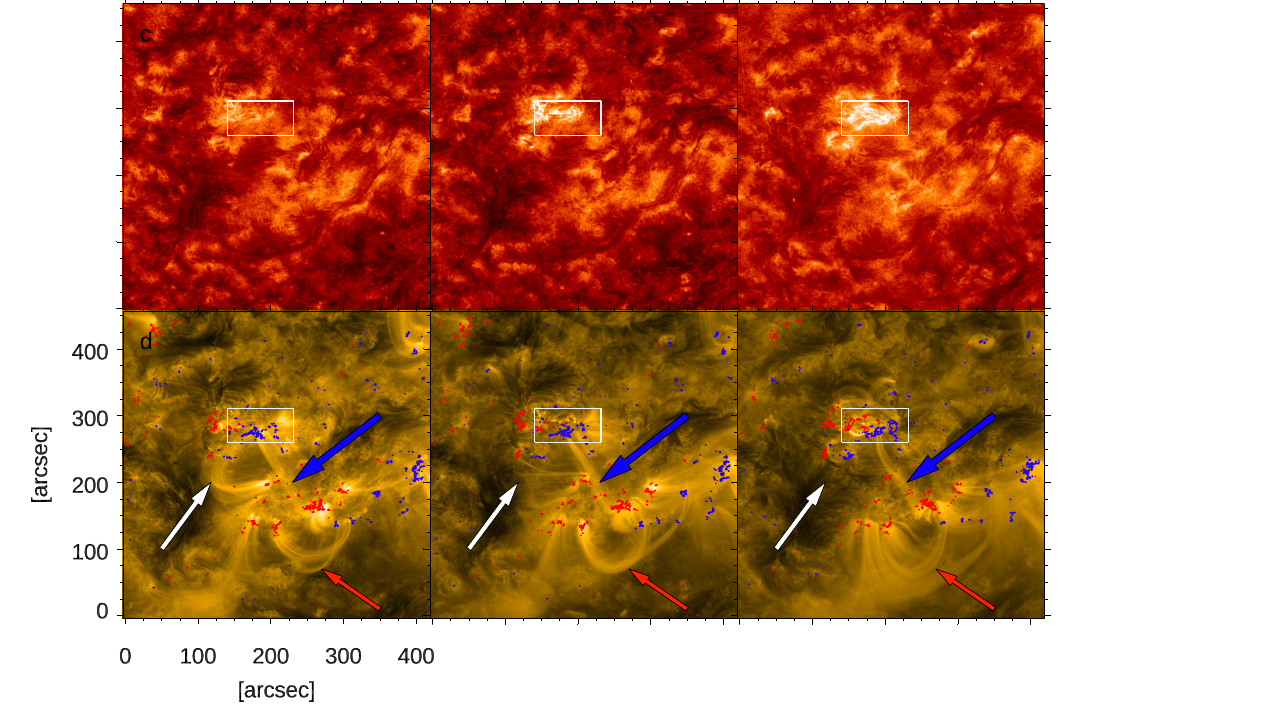}
	\end{adjustwidth}
	\caption{From top to bottom, with letters indicating rows: AR~11462 as seen in the SDO/HMI continuum filtergrams (a) and LOS magnetograms (b), and in the SDO/AIA He~II 30.4~nm (c), and Fe~IX 17.1~nm (d) filtergrams during its evolution at 12:00~UT (left column), 14:00~UT (middle column), 21:00~UT (right column). Red and blue contours at $\pm 300$ G represent negative and positive magnetic flux from SDO/HMI LOS magnetograms. The black (or white) boxes frame the analyzed region. White and black in the magnetograms (row (b)) indicate the positive and negative magnetic flux, respectively. The larger FOV of these images, $\approx 410\arcsec \times 420\arcsec$, is employed to compute the magnetic flux plotted in Figure~\ref{f1}. The insert in panels of rows (a) and (b) displays in more detail the region of interest. Red, blue, and white arrows indicate coronal arcades, as explained in the Sect.~\ref{GO}. Yellow arrow in the middle continuum and magnetogram panels indicates the S-Shape described in the Sect.~\ref{GO}. Here and in the following figures, solar North is at the top, West is to the right. \label{f2}}
\end{figure}

Figure~\ref{f2} shows three representative times of the AR evolution as seen on April 17, 2012 by the instruments aboard SDO. In each panel containing the SDO/HMI photospheric filtergrams and magnetograms (panels a and b), we show an insert of $\approx 87\arcsec \times 50\arcsec$ with the evolving region. 

In addition to IBIS data, flux emergence is also clearly seen in the SDO/HMI magnetograms. They show the growth of magnetic regions by coalescence of smaller scale, same polarity patches, until the formation of a filamentary, counter-S shaped structure (see the yellow arrow in the magnetogram and continuum maps at 14:00 UT of Figure~\ref{f2}), which is reminiscent of flux tube twisting (see for further details Paper I). The initial phase of the flux emergence episode is not covered by IBIS observations, which targeted the evolution of above counter-S shaped structure to form a prominent oval-shaped pore centered at $\approx \left[220\arcsec, 280\arcsec\right]$, from around 14:00 to 21:00~UT.

In Figure~\ref{f2} we also show the analyzed region as observed almost simultaneously by SDO/AIA in the He~II 30.4~nm and Fe~IX 17.1~nm passbands (panels \textit{c} and \textit{d}), which sample the upper chromosphere and the low corona, respectively. We also add the contours of the SDO/HMI LOS magnetic field, which clearly show magnetic concentrations located at the base of coronal arcades. We can easily see that the pore formation occurred in the vicinity of coronal arcades from pre-existing magnetic flux concentrations of an unnamed AR (see Figure~\ref{f2}, panels \textit{d}). These arcades mostly consist of a smaller-scale structure that seems almost parallel to the solar surface (red arrow), and a larger-scale feature protruding into South-West (blue arrow). From 12:00 to 21:00~UT, we can see that the arcades linking AR NOAA~11462 with the unnamed AR (white arrow) disappear.

\begin{figure}[H]
	\centering
	\includegraphics[width=10.5cm, trim=20 0 100 0, clip]{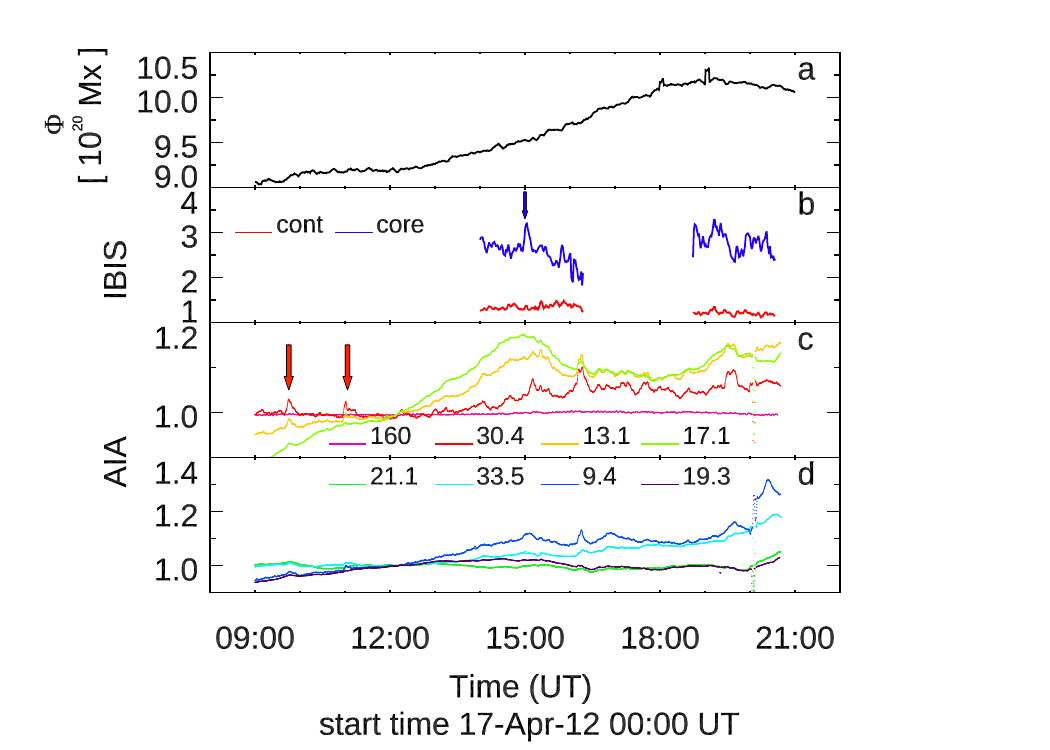}
	\caption{From top to bottom: Variation of the total unsigned magnetic flux in the SDO/HMI magnetograms (panel a)),  total radiative flux in the IBIS Ca~II line data (panel b)) and in the whole FOV extracted from SDO/AIA observations (panels c) and d)) taken on  April 17, 2012 from 09:00 to 21:00~UT, which is the same time interval considered in Figure~\ref{f2}. The values are normalized to the total intensity estimated from the first observation in the analysed time interval. The values for the various passbands (SDO/AIA reported here in nm) and instruments are reported with different colors as specified in the Legend. Blue and red arrows indicate peaks in IBIS Ca~II and SDO/AIA data described in Sect.~\ref{GO}.} \label{f3}
\end{figure}

Figure~\ref{f3} illustrates the evolution of the total intensity in the various sub-fields extracted from the IBIS Ca~II 854.2~nm data, as well as in SDO/AIA observations. The intensities computed from the IBIS Ca~II images, both continuum and line core, generally exhibit an almost constant trend, except for the maximum intensity in the line core (panel b) of Figure~\ref{f3}). Noticeably, the maximum Ca~II line core brightness (see the blue arrow in the panel b of Figure~\ref{f3}) occurs simultaneously to the maximum intensity in the 17.1~nm, 13.1~nm, and 9.4~nm EUV channels, at around 15:00~UT. The 30.4~nm, 17.1~nm, and 13.1~nm total intensities show two peaks (see red arrows in panel c) during the analysed 5 hour time interval preceding the start of the IBIS observations, at the time flux emergence starts (9:00~UT). These peaks are faintly detectable in the 33.5~nm, 19.3~nm, and 9.4~nm EUV filtergrams (panel d). From 12:00~UT the 17.1~nm and 13.1~nm intensities start to gradually increase of about 20\%, while the 30.4~nm intensity increases with a more spiky trend, until it reaches a first maximum peak at around 15:00~UT, co-temporally with the 17.1~nm and 13.1~nm filtergrams. This increase is also visible in the 9.4~nm filtergrams, with a smaller variation with respect to the 17.1~nm and 13.1~nm passbands. Then, a rapid decrease in 17.1~nm and 13.1~nm intensity is followed by an abrupt second maximum at around 16:00~UT, visible also in the 30.4~nm, 33.5~nm and 9.4~nm passbands. The low coronal intensity, except for the 33.5~nm channel, decreases from 16:00 to 19:00~UT, while a new increase with a peak is subsequently detected, during the wrapping of the S-shaped feature after the pore formation, from around 19:00~UT onwards. 

\begin{figure}[H]
	\begin{adjustwidth}{-\extralength}{0cm}
		\centering
		\includegraphics[width=18.5cm, trim=0 130 0 120, clip]{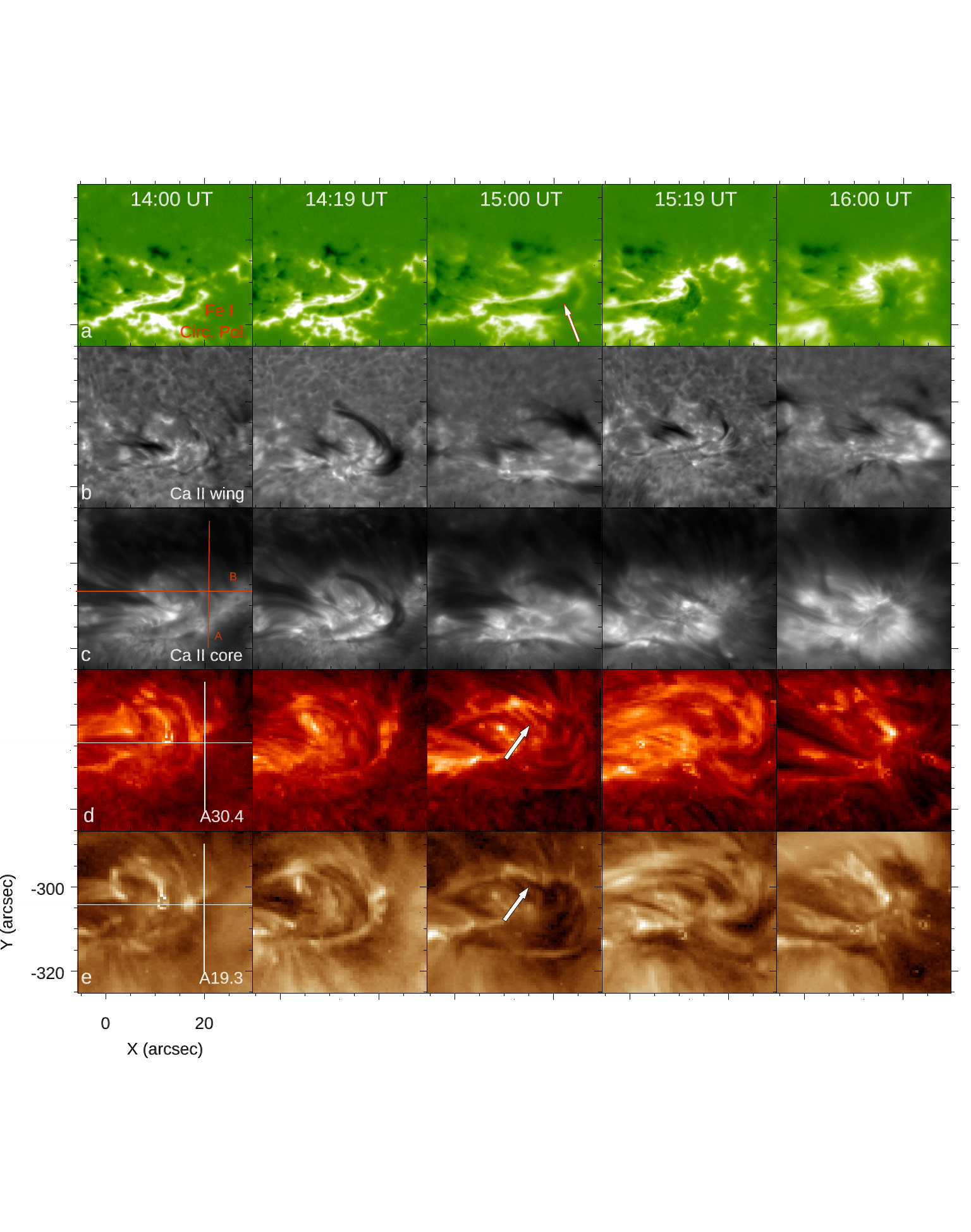}
		\caption{Synoptic view of the event at five representative times during the pore formation from 14:00 to 16:00 UT. From top to bottom:
IBIS Fe I 630.2 nm circular polarization (panels a), IBIS Ca II blue wing and line core (panels b and c), SDO/AIA 30.4 nm (panels d), SDO/AIA 19.3 nm (panels e) observations. The white and red segments overplotted to IBIS Ca II line core and SDO/AIA 30.4 nm, 19.3 nm indicate the horizontal and vertical slits used to make the time-slice plots shown in Figure\ref{f5} and Figure\ref{f9}. Red and white arrows indicate structures described in the text. A movie of the IBIS data (Ca II wing and core) showing the entire FOV observed in the time interval between 13:58 UT and 15:00 UT is available online. \label{f4}}
	\end{adjustwidth}
\end{figure}

\begin{figure}[H]
	\begin{adjustwidth}{-\extralength}{0cm}
	\centering
		\includegraphics[width=15.5cm, clip, trim=0 330 100 190]{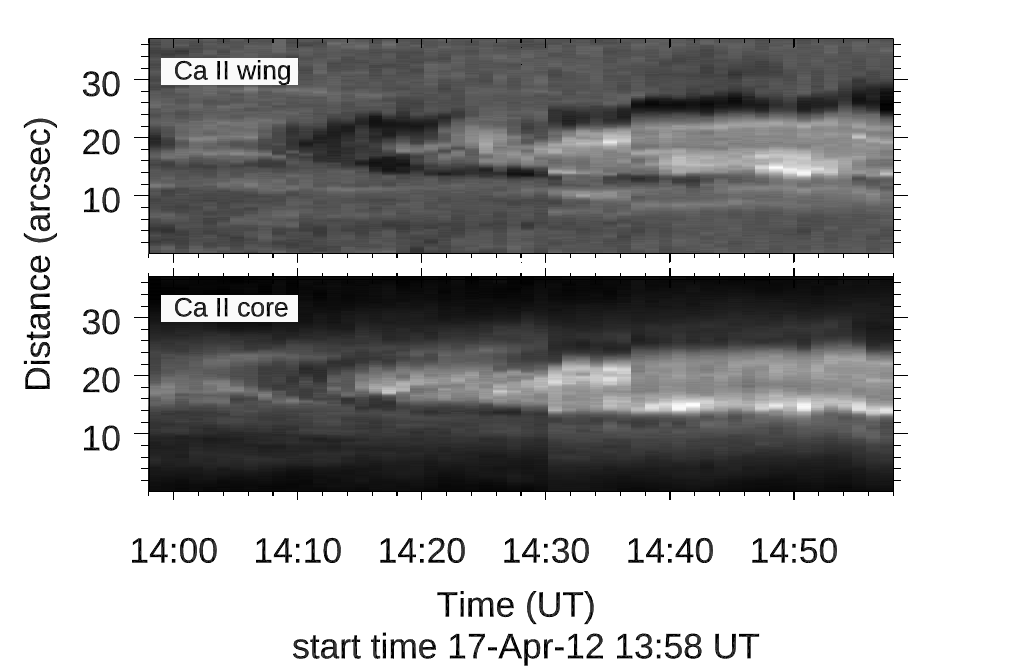}
		\includegraphics[width=15.5cm, clip, trim=0 300 100 190]{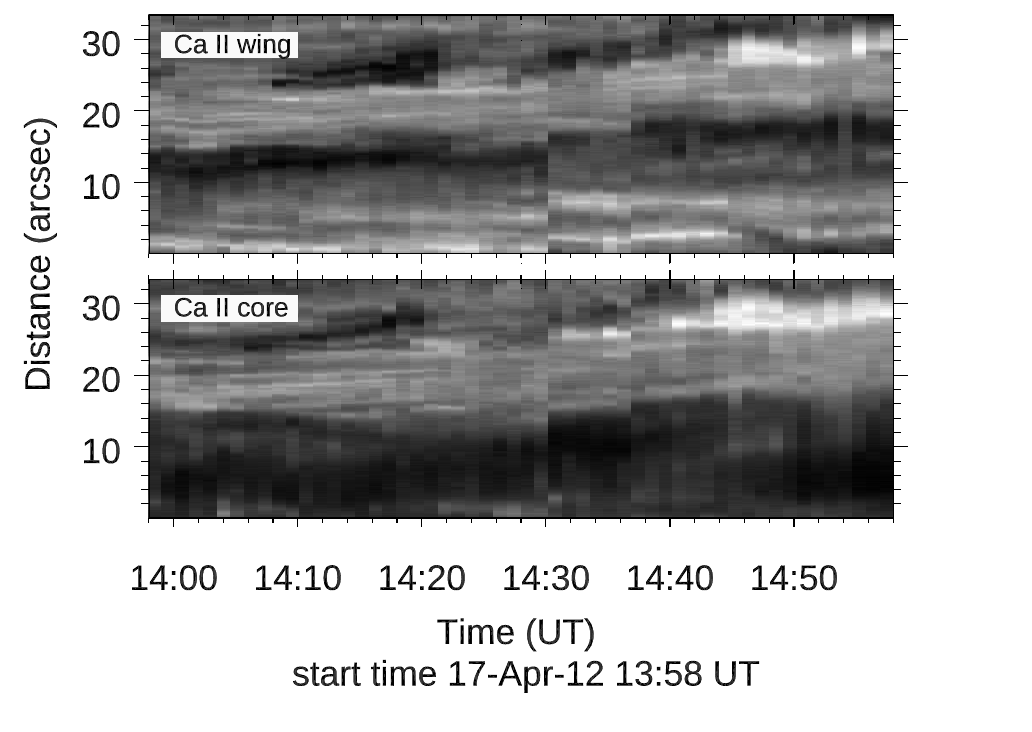}\\
	\caption{Time-slice plots relevant to the slits indicated with red segments in Figure~\ref{f4}: top panels, slit A (vertical); bottom panels, slit B (horizontal). These slice-plots allows us following the evolution of the surges seen in the IBIS Ca~II line wing and line core observations. \label{f5}}
	\end{adjustwidth}
\end{figure}

\subsection{IBIS chromospheric observation}

The evolution of the analyzed region deduced from the photosperic IBIS and SDO/HMI observations taken at the Fe~I 617.3~nm line, which was studied in detail in Paper~I, includes i) the appearance of elongated granules and bright points in the vicinity of the stronger flux patches of the EFR; ii) counter-streaming of mixed polarity small-scale magnetic flux concentrations; and iii) motion of the formed elongated magnetic patch to the West and rotation in the counter-clockwise direction. We present in Figure~\ref{f4} panels a the evolution of the integrated circular polarization signals in the Fe~I 630.2 nm line revealing ribbon-like features with positive polarity along the filamentary S-shaped structure and patches of opposite polarity inside the ribbon-like features.

Figure~\ref{f4} also displays an overview at high spatial resolution of the chromospheric morphology of the evolving region, occurring during the early stages of the pore formation as seen in the line wing (panels )) and line core (panels c) IBIS Ca~II data. At the beginning, at X=[-5\arcsec,10\arcsec] and Y=[-310\arcsec,-305\arcsec], we see the presence of an AFS connecting the opposite polarities of the EFR, at the edge of the evolving region and in the vicinity of the pre-existing positive polarity patches seen in the polarization maps (panels a). In the IBIS observations we see only a portion of the AFS due to the limited IBIS FOV. The entire feature is clearly observable, e.g., in the SDO/AIA 30.4~nm filtergrams shown in Figure~\ref{f7}.


At the edges of the AFS, we find surges, which are cool and dense chromospheric plasma ejections \cite{Daniel:21}, consisting of small-scale thread-like features \cite{Zhen2016}. In particular, they appear southward and westward of the AFS. Very interestingly, the surges are seen to occur following the counter-clockwise direction as time evolves. Both the line wing and line core observations (panels b and c) also show brightening events at the edge of these structures. The features on the South and West sides of the region exhibit rather similar characteristics in terms of size and propagation. 
Their speed slightly increases in time 
while their occurrence decreases over the later stages of the pore formation.

Similarly to what seen in the polarization maps, the edges of the evolving region, which appears as a compact bundle in core of the Ca~II line, expand and move westward. In the horizontal plan, we estimate an average value of the diverging motion velocity $\approx 1.5 \,\mathrm{km/s}$. This can be noticed from the time-slice plots displayed in Figure~\ref{f5}, relevant to the slit A and B represented with red segments in Figure~\ref{f4} (panels c, first column).
 
Interestingly, the surges seen in the IBIS Ca~II line wing data set in while the magnetic flux is still increasing in the upper photosphere. This suggests that the magnetic reconnection that supplies the energy of the surges occurred above the IBIS Fe~I 630.2~nm line formation heights, specifically in the low chromosphere, where the new emerged field interacts with the pre-existing environment. Opposite patches that get into contact are indeed observed in the polarization maps (Figure~\ref{f4}, panels a).

\begin{figure}[H]
		\centering
		\includegraphics[scale=0.425, clip, trim= 50 70 120 10]{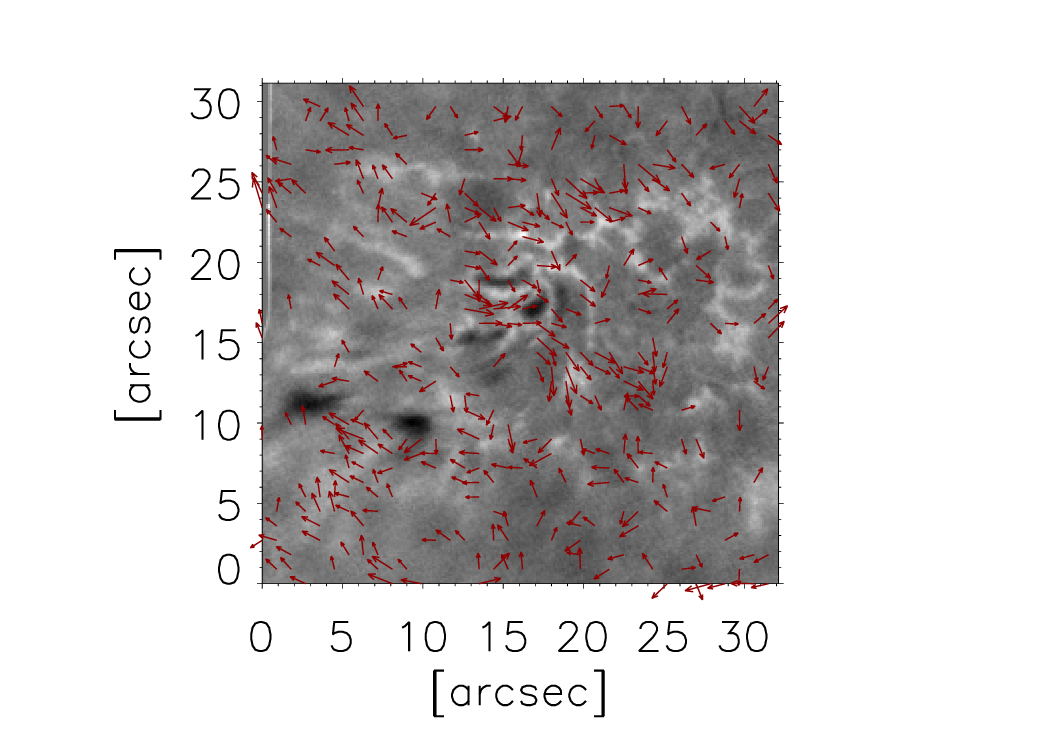}
		\includegraphics[scale=0.425, clip, trim= 120 70 120 10]{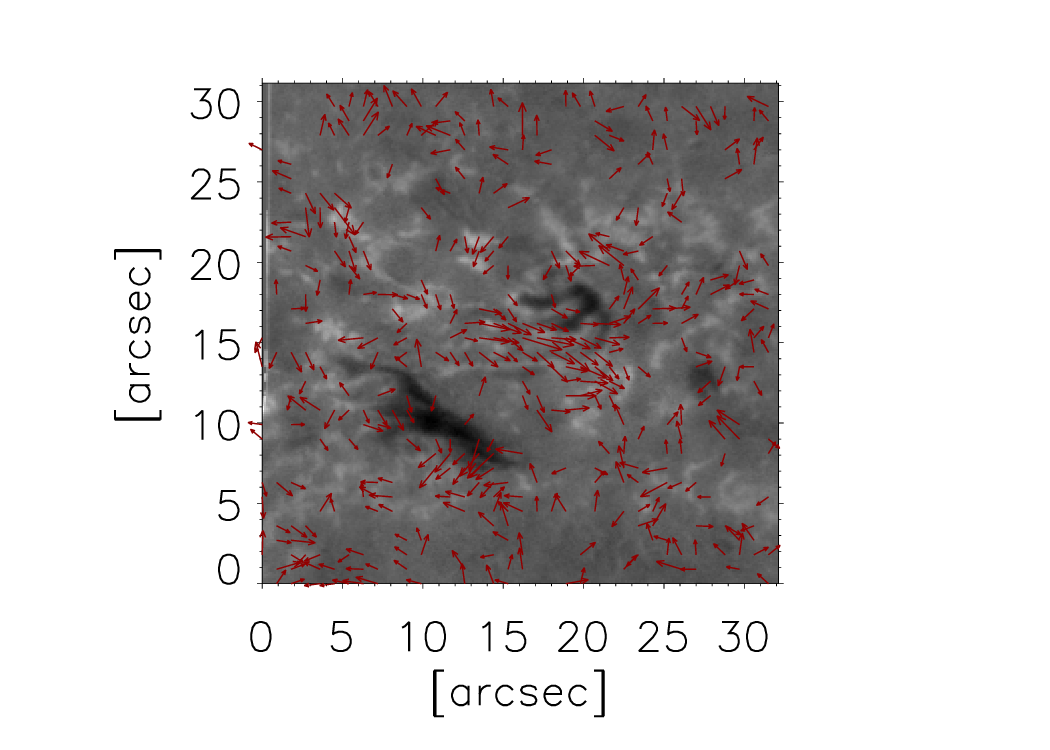}
		\includegraphics[scale=0.425, clip, trim= 120 70 120 10]{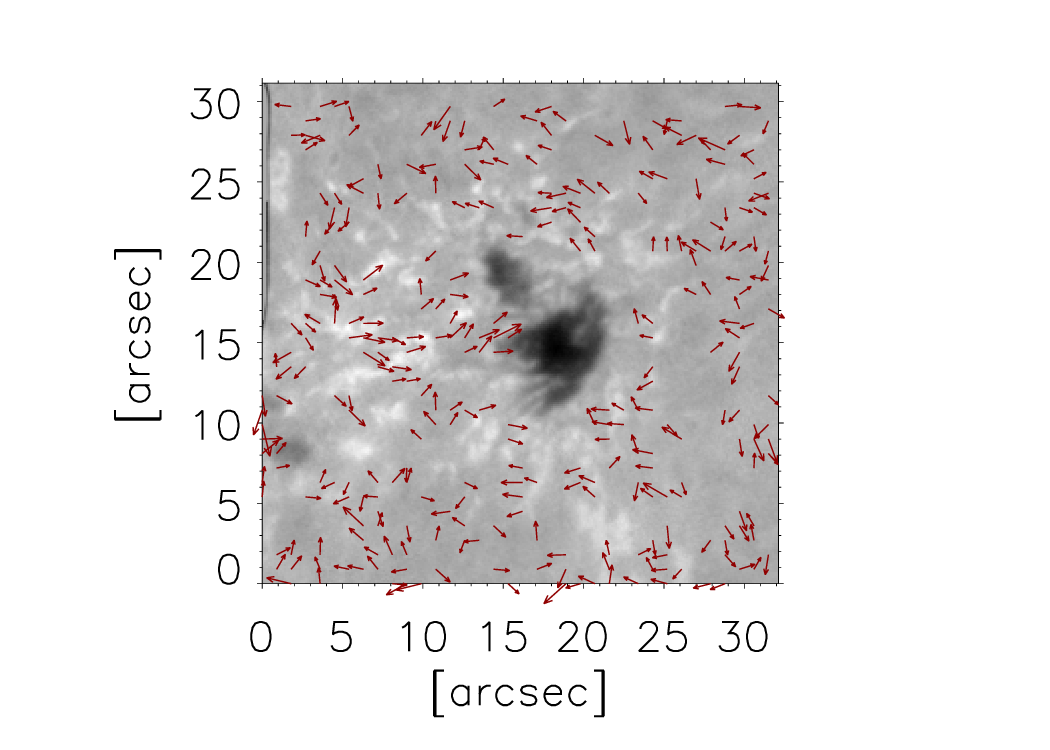}
		\includegraphics[scale=0.425, clip, trim= 50 70 120 10]{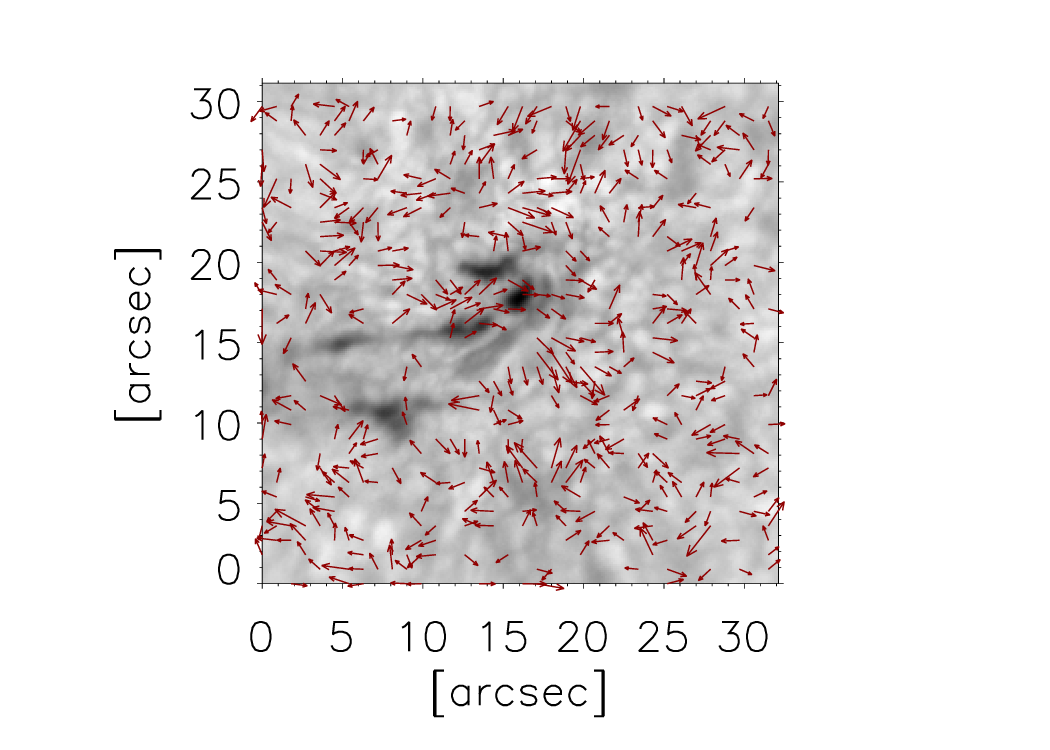}
		\includegraphics[scale=0.425, clip, trim= 120 70 120 10]{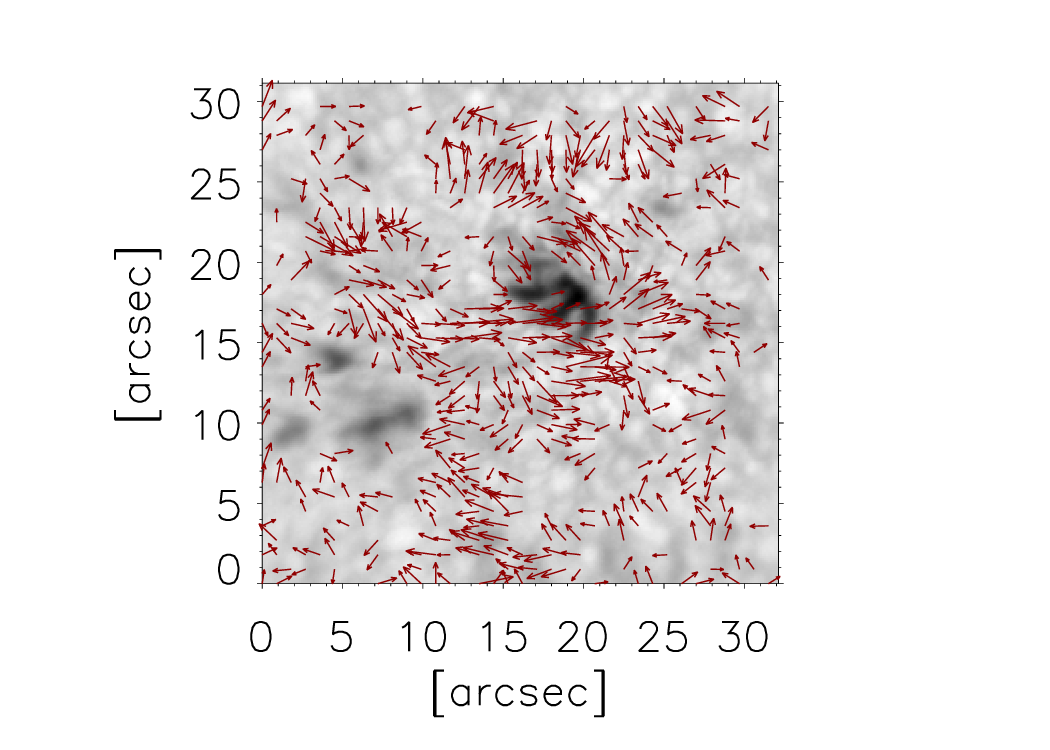}
		\includegraphics[scale=0.425, clip, trim= 120 70 120 10]{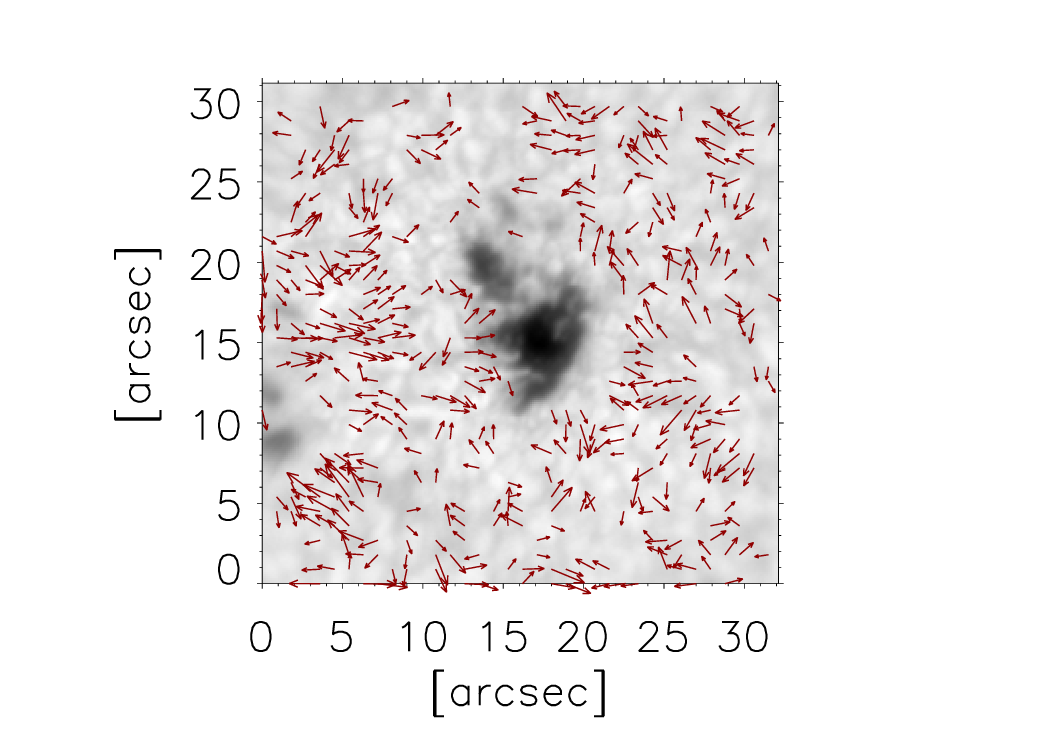}
		\includegraphics[scale=0.425, clip, trim= 50 0 120 10]{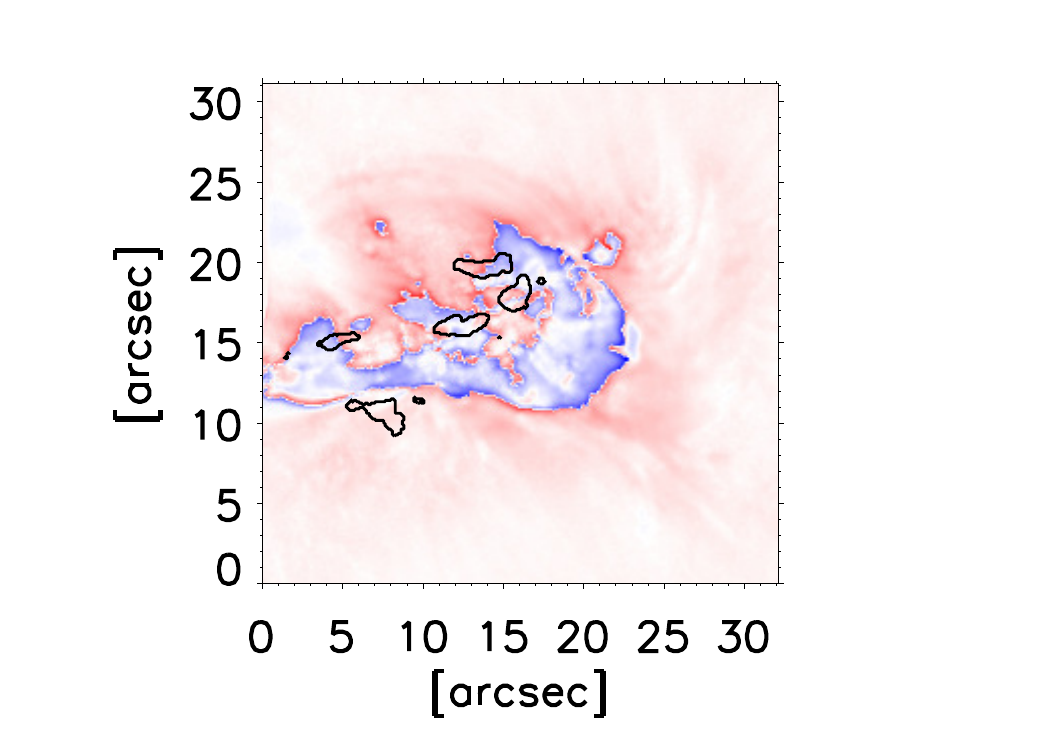}
		\includegraphics[scale=0.425, clip, trim= 120 0 120 10]{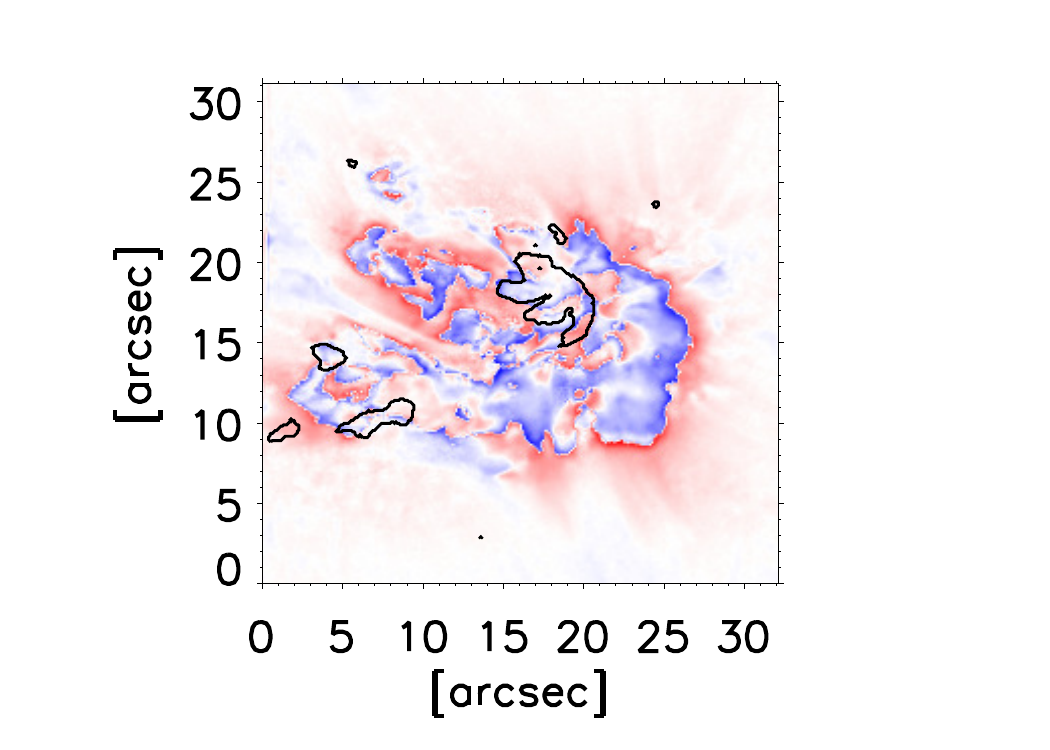}
		\includegraphics[scale=0.425, clip, trim= 120 0 120 10]{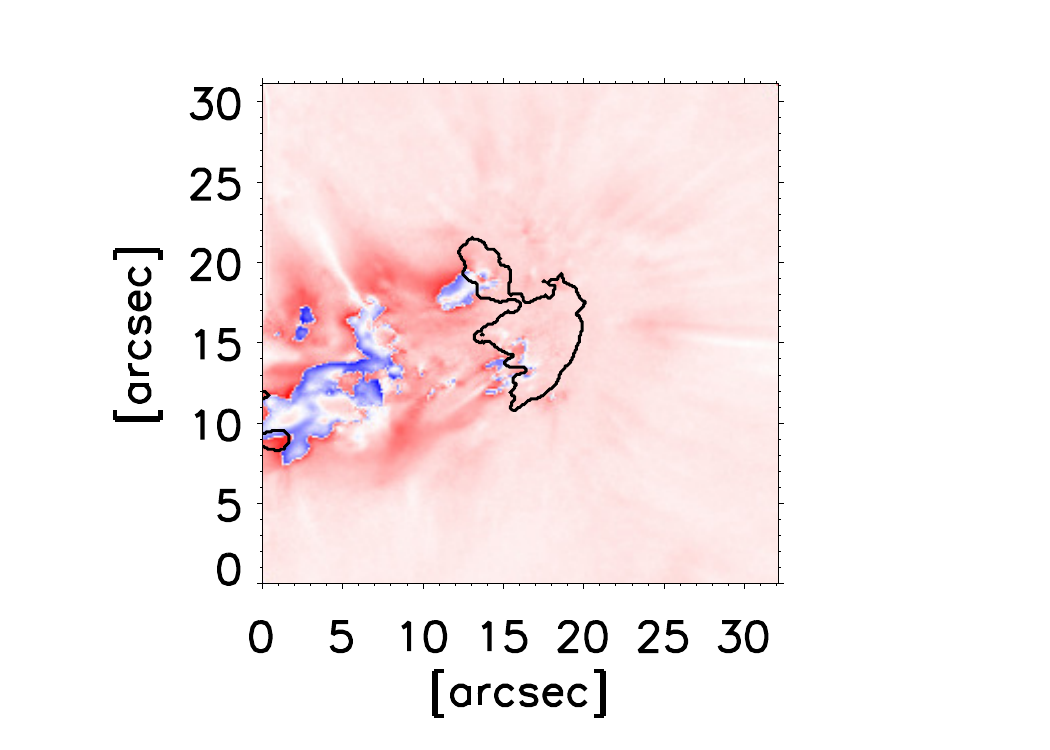}
	\caption{Sample of the IBIS observations at three stages: 15:08~UT (left column), 16:08~UT (middle column), and 19:29~UT (right column). From top to bottom: observations at Ca~II line core (top panels), Fe~I continuum (middle panels) and Ca~II LOS velocity (bottom panels). Overplotted the $\mathrm{v}_{H}$ maps (top and middle panels), representing the horizontal plasma velocity fields in the evolving region derived from the IBIS Ca~II 854.2~nm (top panels) and Fe~I 630.2~nm (middle panels) observations. The intensity background in the $\mathrm{v}_{H}$ maps shows either the average image of the representative series or the evolving region. The contour shows the location of the evolving structure singled out in the continuum data, as defined by the intensity ($I$) threshold $I = 0.9 \, I_{QS}$ where $I_{QS}$ is the mean intensity of quiet Sun in the studied region. Red/Blue colors indicate areas with downflows/upflows, respectively, saturated at $\pm 20 \,\mathrm{km/s}$. \label{f6}}.
\end{figure}

Figure~\ref{f6} shows Ca~II line core intensity maps, as well as $\mathrm{v}_{H}$ maps derived from the chromospheric measurements and from the Fe~I 630.2~nm photospheric data, at three representative stages during the IBIS observations (from 15:08 to 19:29~UT). The concept of the figure is similar to what already presented in Paper I with the difference of the photospheric line used (Fe I 630.2~nm instead of Fe I 617.3~nm) and including the Ca II spectroscopic observations.

Before the appearance of the pore, until 16:08~UT (Figure~\ref{f6}, top panel), some fibrillar structures are visible in the Ca II core. After its formation, the pore appears dark and the radiative emission of the plage surrounding the evolving feature overcompensates the emission deficit due to the forming pore during its entire evolution.

The chromospheric $\mathrm{v}_{H}$ maps (Figure~\ref{f6}, top panels) display larger-scale flow patterns, with lower velocities than those derived from similar photospheric data (Figure~\ref{f6}, middle panels), being on average $<0.3 \,\mathrm{km/s}$ instead of about $1 \,\mathrm{km/s}$. The maximum $\mathrm{v}_{H}$ value measured in the chromosphere does not vary significantly during the pore formation. 

Conversely, photospheric $\mathrm{v}_{H}$ maps show a more significant enhancements of translational and rotational motions in the counter-clockwise direction of the positive polarity at around 16:08~UT. These motions drag the evolving feature to push and slide towards the neighboring pre-existing flux region, with the same polarity, located in the southward direction, as already noticed from the SDO/HMI in Paper~I (Figure 4), where we also detected opposite rotation of the two main polarity patches of the EFR. These motions likely increase the shear and build-up of magnetic energy in the region, which is then released during several brightening events, surges, and jets observed higher in the atmosphere.

The $\mathrm{v}_{LOS}$ maps (Figure~\ref{f6}, bottom panel) derived from the Ca~II chromospheric data exhibit small-scale patches of upflows and downflows close by each other. Supersonic plasma flows near the edges of the evolving region, with velocities up to $20 \,\mathrm{km/s}$. Some of the flows are observed near areas of the evolving region, where the core of the spectral line goes into emission and the Doppler shift is not always measurable.

A comparison with Figure~\ref{f4} (panels a, b, and c) suggests that IBIS chromospheric data show localized brightening events where the photospheric plasma motions are converging, in the contact region between the merging positive polarity patches from the EFR and the positive polarity structure aligned to the evolving region (see the red arrow in Figure~\ref{f4}, panel a at 15:00~UT). In this region, minority opposite polarity patches are found.



\subsection{Response to pore formation in SDO/AIA channels, up to coronal levels}
\label{porefor_salvo}

The comparison between IBIS chromospheric observations and SDO/AIA images at 30.4~nm and 19.3~nm in Figure~\ref{f4} (panels d and e) reveals that the footpoints of the small-scale emerging loop systems ensued from the EFR are anchored in the emerging flux concentrations and move as these patches move and rotate. The shape of the emerging loops is suggestive of a twisted field configuration, also reminiscent of the translational and rotational motions of the evolving feature observed in the photosphere.

In addition to the coronal environment at the pore formation site, SDO/AIA images show the dynamical creation and rearrangement of existing loop systems, with associated brightening and plasma ejections (e.g., the site indicated by the white, blue and red arrows in Figure\ref{f2} d). It is worth noting that a canopy-like structure forms around the pore during about 6 hours, starting from 14:30~UT. We remember that also the arcades protruding into South-West (blue arrow in Figure~\ref{f2} panels d) change with time. 

Both the SDO/AIA 30.4~nm and IBIS Ca~II data display bright patches southward and eastward of the forming pore, with clear counterparts in the 19.3~nm observations. The emission in these EUV passbands is confined largely to loops anchored in and around the evolving feature. The atmosphere therein is likely site of enhanced heating due to the formation of strong electric current layers.

\begin{figure}[H]
	\begin{adjustwidth}{-\extralength}{0cm}
		\centering
		\includegraphics[width=18.5cm, trim=0 320 0 0, clip]{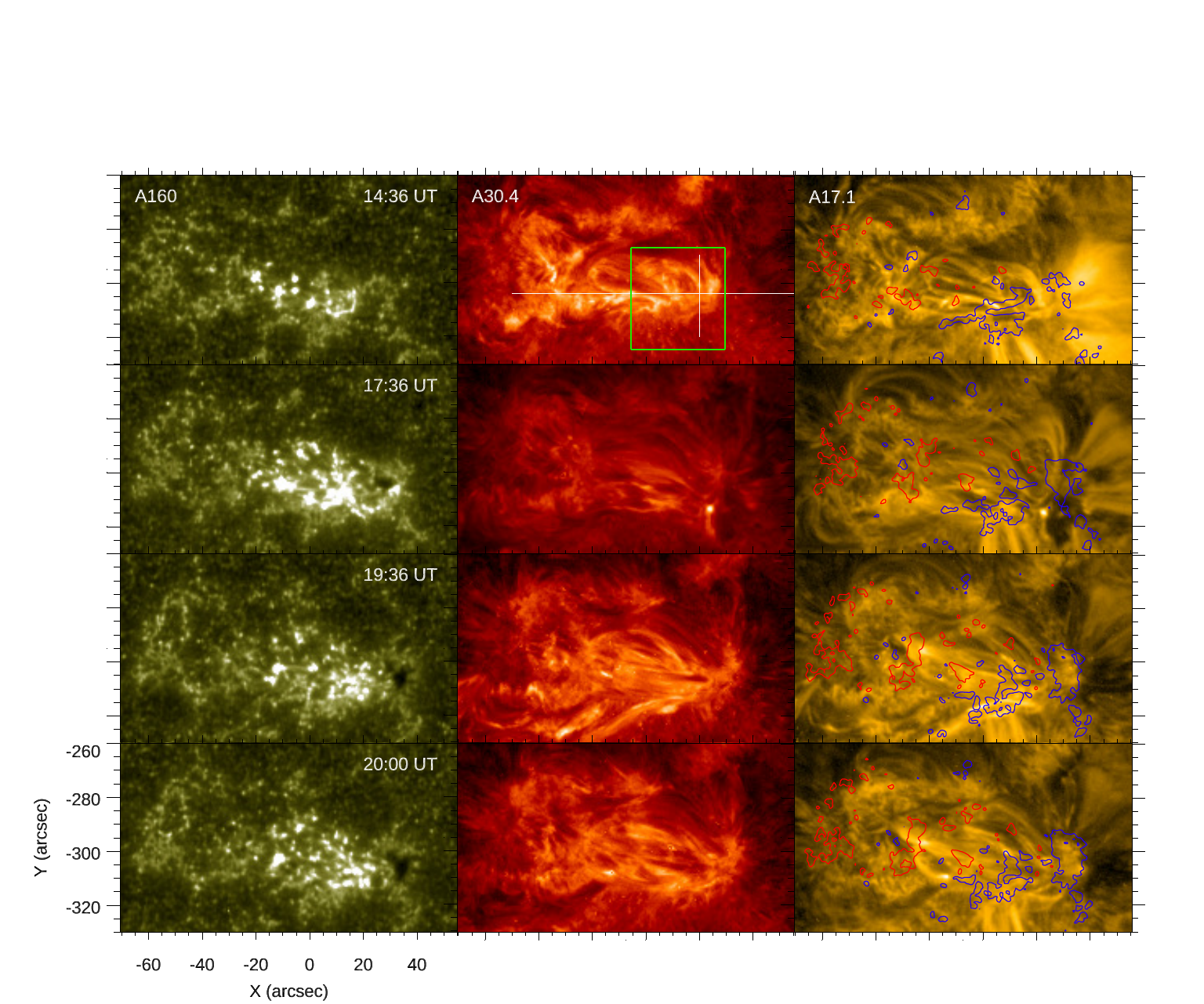}
\includegraphics[width=18.5cm, trim=0 0 0 700, clip]{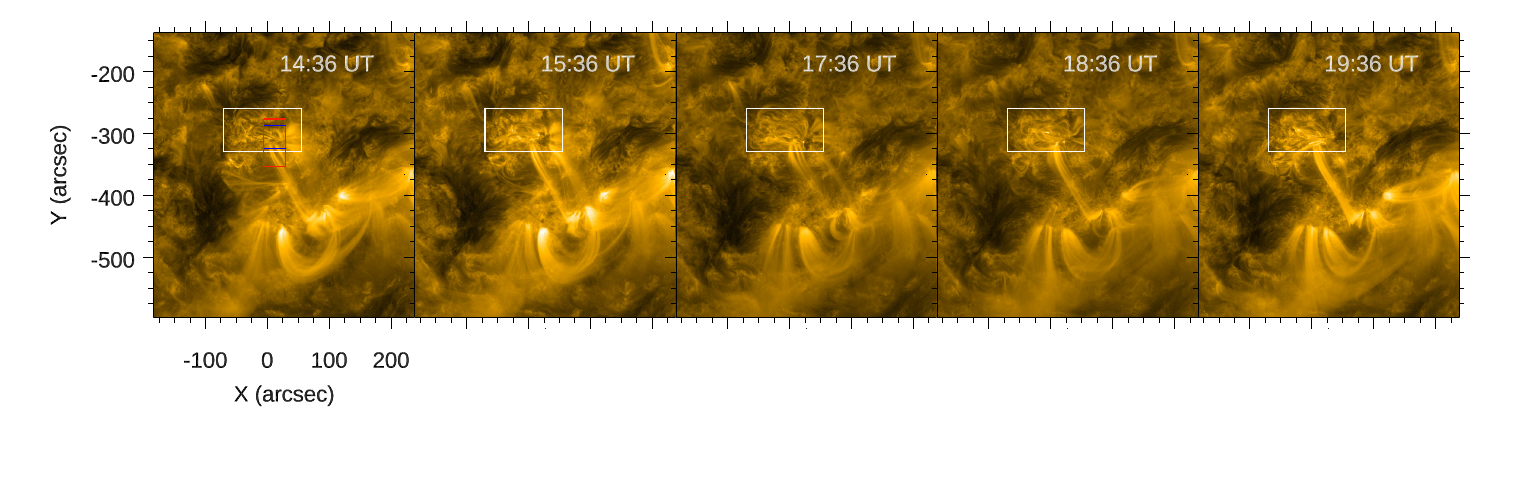}		
		\caption{Top panels, from left to right: SDO/AIA  160~nm, 30.4~nm and 17.1~nm intensity maps to follow the evolution of the ROI during the pore formation from 14:36 (top raw) to 20:00 UT (bottom row). The blue dashed box in the middle top panel indicate the ROI shown in Figure~\ref{f6} and ~\ref{f8}. Bottom panels from left to right: SDO/AIA filtergrams at 17.1~nm from 14:36 to 20:00~UT, covering the extended FOV of Figure~\ref{f2} that includes the studied region. The white box indicates the ROI shown in the top panels, while the red box frames the full FOV of the IBIS observations for comparison}. \label{f7}
	\end{adjustwidth}
\end{figure}

\begin{figure}[H]
	\begin{adjustwidth}{-\extralength}{0cm}
	\centering
	\includegraphics[width=18.5cm, trim=20 320 0 100, clip]{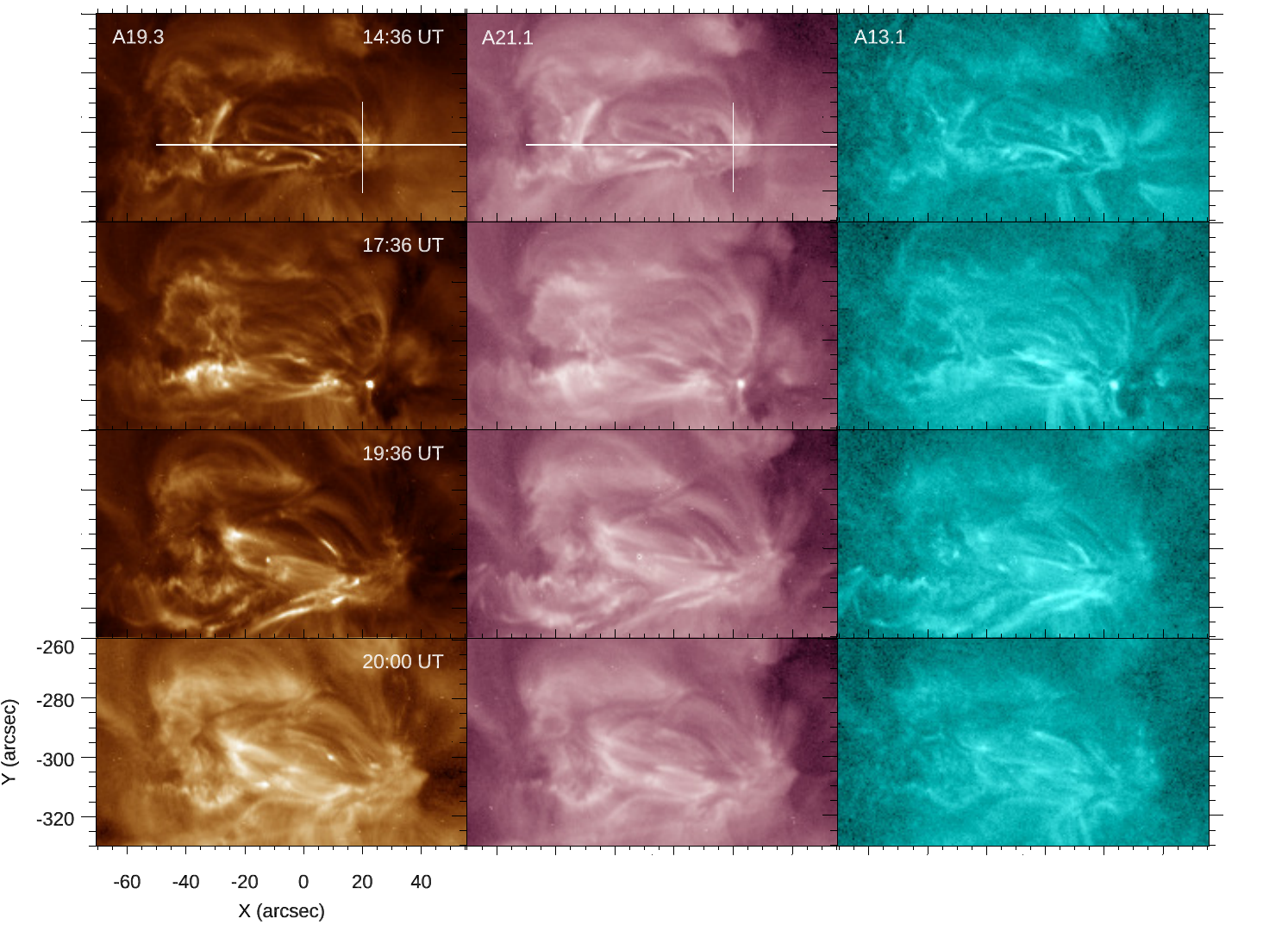}
	\caption{From left to right: SDO/AIA 19.3 nm, 21.1 nm and 13.1 nm intensity maps to follow the evolution of the ROI
during the same time interval of Figure~\ref{f6}.} \label{f8}
	\end{adjustwidth}
\end{figure}

In Figures~\ref{f7} and~\ref{f8} we show a synoptic view, from the upper photosphere to the corona, of the evolution of the region where the pore later will appear (at 17:36~UT) for more than five hours. Figure~\ref{f7} (top panel) displays 160~nm, 30.4~nm, and 17.1~nm filtergrams co-aligned, with contours from SDO/HMI LOS magnetic field over-imposed to the 17.1~nm passband. Similarly, in Figure~\ref{f8} we display simultaneous 19.3~nm, 21.1~nm, 13.1~nm filtergrams. The subFOV used in these images is marked in the SDO/AIA 17.1~nm images in Figure~\ref{f7} (bottom panel).

The magnetic dynamics in the photosphere affects the upper atmospheric levels. Initially, at 14:36~UT, the 160~nm bandpass is characterized by some dot-like brightening events localized all around the S-shaped feature visible in the nearly-simultaneous SDO/HMI magnetograms (e.g., the insert in Figure~\ref{f2}, top-middle panels) as well as in the IBIS observations at 14:00-14:19~UT. Note that the SDO/AIA 160~nm bandpass refers to the upper photosphere, even if there is a significant contribution from C IV lines originating in the transition region. It is possible to note the effect of the displacement of the evolving pore in the upper photosphere, which is probably due to the cancellation/merging of the magnetic flux during the drift motion.  

Also the upper chromospheric 30.4~nm display localized and often elongated brightening events, similar to bright fibrils, during the entire analyzed time interval, not only in the region zoomed in Figure~\ref{f4} but also in the entire EFR. The two polarities appear to be connected by an AFS at X=[-40\arcsec,20\arcsec] and Y=[-320\arcsec,-280\arcsec] in these 30.4~nm filtergrams, with some knots with enhanced emission at the footpoints (see Figure~\ref{f7} middle panels). This feature shows its major evolution at around 19:36~UT, when a brightening is also seen in the IBIS data, while the photospheric plasma flows in the newly formed pore evolve (see right panels in Figure~\ref{f6}). 

A highly dynamic evolution is also found in the 17.1~nm filtergrams (see top-right panels of Figure~\ref{f7}). A large darkening is observed south to the region hosting the pore formation at around 17:36~UT. Data taken at around 19:36~UT show a localized brightening in the region of the AFS. 

Brightening knots are also visible at larger scale in the 17.1~nm filtergrams (see Figure~\ref{f7} bottom panels). These data illustrate that the connection with the unnamed AR undergoes a significant rearrangement and subsequent enhancement. We also see that the pre-existing large-scale loop system, involving the central part of AR NOAA~11462 and protruding to South, displays intense brightening in correspondence to the forming pore (18:36~UT). Afterwards, the connection remains strong, exhibiting a ribbon-like loop system with enhanced emission.

Along the axis of the S-shaped feature, intensity enhancements are mostly visible in 30.4~nm, as well as in the 21.1~nm, and 13.1~nm filtergrams (Figure~\ref{f8}, left and middle panels), which can be attributed to interaction with the ambient field caused by the positive polarity migration and the subsequent pore formation. However, these variations are not visible in 17.1~nm images. 

The AFS displays some dynamics, mostly in the EUV 17.1~nm data, with events of both brightening and loop evolution. SDO/AIA data in other channels (30.4~nm, 21.1~nm, and 13.1~nm, not shown in Figure~\ref{f4}) show the same arch-shaped feature that opens up / expands towards the future pore (e.g., white arrows in Figure~\ref{f4} panels d and e). 

During all of the stages of the pore formation, SDO/AIA 30.4~nm observations show the above AFS anchored in the evolving structure and near negative polarity patch, with clear counterpart in the 17.1~nm data. This counterpart is clearly seen in the 19.3~nm data as well, while it is less evident in the hotter SDO/AIA diagnostics.

Recurrent intense brightening with counterpart in all (if not, in many) SDO/AIA passbands are identified at the edges of the EFR. Jet activity is also observed at coronal level. In particular, SDO/AIA observations at 19.3~nm and 21.1~nm show coronal jetlets, e.g. at around 16:45~UT (e.g. arrows in Figure~\ref{f9}, see below), which are spatially separated from chromospheric surges, and initiated at the late stages of the chromospheric surges observed in the IBIS data. However, brightening cospatial to the EFR are seen since 14:36~UT in all SDO/AIA EUV channels, when a bright blob is found with the base almost at the same site of the first surge.

Interestingly, a dot-like very intense brightening is seen in all the EUV passbands of SDO/AIA simultaneously at 17:36~UT South to the forming pore, which suggests the occurrence of a micro-flaring event.
	
In order to explore these processes, we analysed time-slice plots obtained in select SDO/AIA EUV channel data at two different locations in the evolving region, marked with white lines in Figures~\ref{f7} and \ref{f8}. These refer to two relevant positions in the EFR, namely the horizontal axis,  and another vertical axis at the western periphery of the EFR, with different lengths.

Figure~\ref{f9} shows the time-slice plots at the two above locations derived from the SDO/AIA 30.4~nm, 19.3~nm, and 21.1~nm filtergrams. The analysis of these time-slice plots indicates that several small-scale brightenings are observed above the EFR, moving along the threads of the small-scale loops as small opposite-direction flow patches. 

The slit along the horizontal direction (Figure~\ref{f9}, bottom panels) clearly displays the expansion of the AFS. The computed expansion rate is about $2 \,\mathrm{km\, s^{-1}}$. Moreover, from 16:00~UT onward, we find the occurrence of several ejections that are visible as threads to the West, where the emerging field meet the pre-existing ambient field. The vertical slit passing at the edge of the EFR (Figure~\ref{f9}, top panels) reveals the arrival of the AFS at around 16:30~UT, with the replacement of the loop arcade, causing the enhanced brightness, with the dark region of the AFS and the threads above noticed, due to plasma ejections.

\begin{figure}[H]
	\begin{adjustwidth}{-\extralength}{0cm}
		\centering
		\includegraphics[width=12.4cm, trim=10 275 160 185, clip]{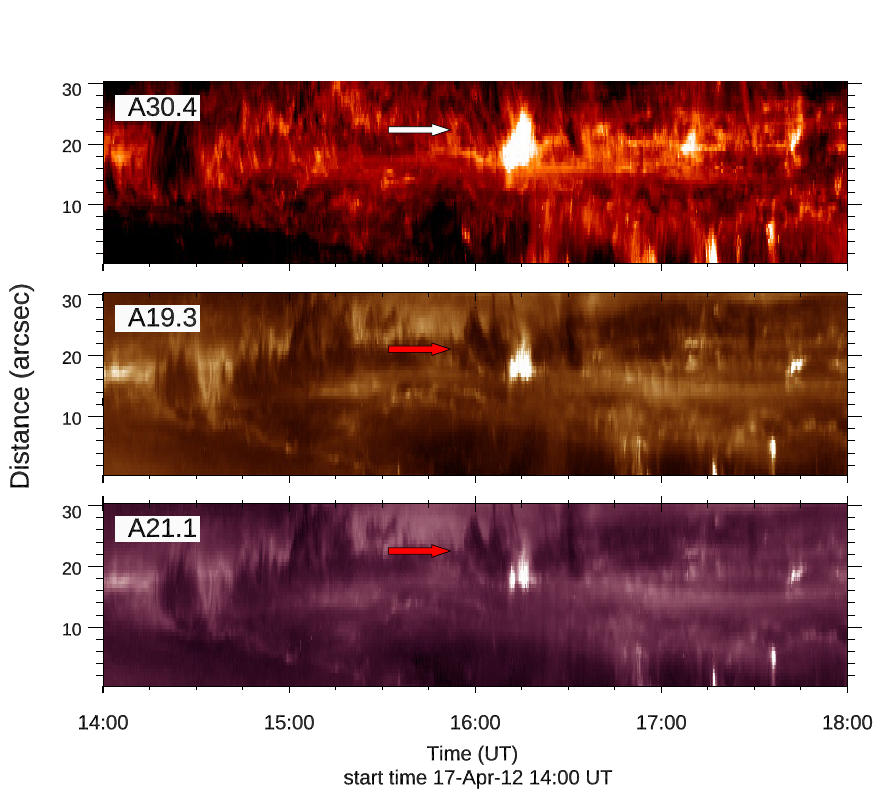}
		\includegraphics[width=12.4cm, trim=10 225 160 185, clip]{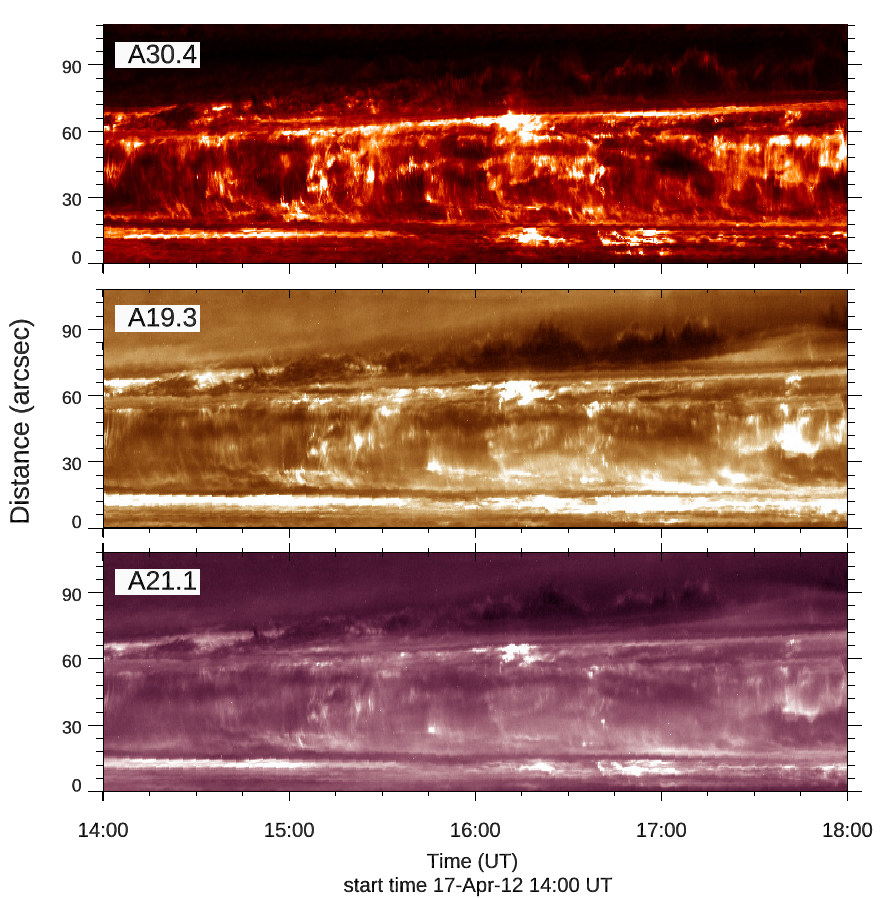}
		\caption{Time-slice plots for SDO/AIA 30.4~nm, 19.3~nm, and 21.1~nm observations: horizontal slit (bottom panels) and vertical slit (top panels). Slits are indicated with white segments in Figure~\ref{f4}, Figure~\ref{f7} and Figure~\ref{f8}.} Red and white arrows indicate the location of a jet. \label{f9}
	\end{adjustwidth}
\end{figure}

\newpage

\section{Discussion}


In recent years, several investigations benefited from availability of multi-wavelength observations to reveal the atmospheric response relative to the appearance of small-scale EFRs, using high-resolution data sets. However, very few works based on such observations concerned the formation of a relatively large-scale magnetic feature as reported in this study. 

The observations here analysed show that the initial stages of the pore formation are characterized by a steady flux increase, which takes place in a pre-existing magnetic environment with small- and large-scale overlying arcades (Figure~\ref{f2} d). The EFR producing the pore is accompanied by energy release phenomena in the chromosphere and upper atmosphere. The emerging magnetic flux interacts with ambient fields already present at the emergence site, by perturbing local conditions with processes that involve magnetic reconnection and conversion of magnetic energy stored in field lines into heat and kinetic energy and manifested as brightenings and jets. 

Brightening events manifest in all the SDO/AIA observations as dot-like features that elongate along the emergence axis after the first flux emergence episode in the region, i.e., quicker than reported by \cite{Tarr:14}. They mark the interaction of the newly emerging flux with the pre-existing coronal field. This interaction appears in the EUV lightcurves in Figure~\ref{f3} as the brightness enhancements seen between 14:00 and 15:30~UT.

While the footpoints separation of the EFR increases, due to the diverging motion, and emerging field lines move upward, they push the overlying arcades, which start to reconnect and produce new loop configurations at small and large scales. Furthermore, the leading polarity patches of the EFR slip along the contact region with the northern footpoint of the pre-existing large-scale coronal arcade. Both brightening events and the modification and/or creation of the pre-existing coronal loops above the emergence site have been already reported by \cite{Zheng:20} as a result of pore formation in the center of an active region. 

To further study the topology of the coronal field at the site of the EFR, and their interaction with the ambient field, we carried out magnetic field extrapolations based on SDO/HMI SHARP magnetograms taken at given times during the pore formation. This allows us to describe the global topology of the field in the evolving region, as well as the overall orientation of the field lines above the EFR and the evolving counter-S shaped feature.  

At the early stages of the pore formation, the coronal loops observed in the EUV images reveal a non-stressed magnetic configuration, with no sigmoid structure and loops oriented parallel to each other, as seen until 17:36~UT in Figure~\ref{f7} (bottom panels). This suggests a low-energy state of the evolving region at the beginning of the pore formation, which might be rather well represented by the performed extrapolations. 

\begin{figure}[H]
	\centering
		\includegraphics[width=7.5cm, trim=100 0 65 0, clip, angle=-90]{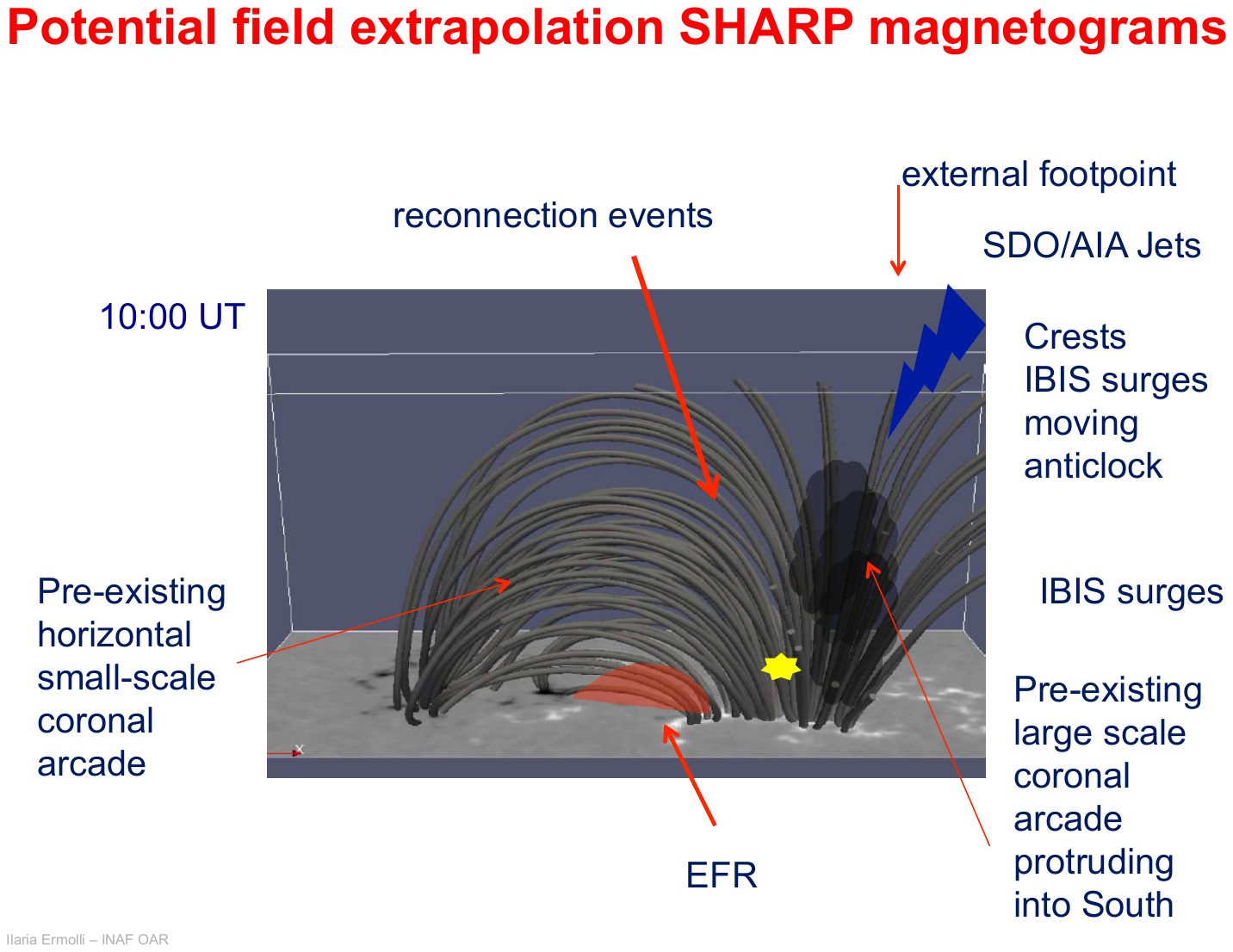}%
		\caption{Cartoon of the observed phenomena, such as EFR, IBIS surges and the SDO/AIA jets on the potential field extrapolation (side view), performed with the method by \cite{Alissandrakis:1981} using SDO/HMI SHARP magnetograms acquired at 10:00 UT, representative of the photospheric boundary conditions. The yellow star represents the reconnection site. 
        \label{f13}}
\end{figure}

Figure~\ref{f13} reports a cartoon, based on a potential magnetic field extrapolation, illustrating the observed phenomena in the IBIS and SDO/AIA observations along with the global topology of the field in the evolving region and overall orientation of the field lines above the EFR at 10:00~UT. 
The large-scale magnetic field connectivity before the EFR suggests a bipolar configuration throughout all the analysed times. The potential magnetic field extrapolations at the early stages show the contact region (yellow asterisk in  Figure~\ref{f13}) between the edge of the EFR and the pre-existing patches of positive polarity. The surges observed in the IBIS chromospheric data (black clouds in Figure~\ref{f13}) are triggered at this contact region, at the boundary between the new and pre-existing magnetic domains.

The chromospheric surges are seen almost co-spatial with the field lines produced by the convergence of arcades overlying the EFR and pre-existing loops (Figure~\ref{f2} d). Thus, they could represent plasma ejecta, accelerated through reconnection, propagating in the convergence region of these arcades at the contact site of the above arcades with field lines of minority negative polarity patches (observed at the time the counter-S shaped feature is formed). The emerging field lines seem not to participate in the surges, but they indirectly seem to trigger the surges by pushing the ambient field together and forcing magnetic  reconnection.
Moreover, the presence of the outer pre-existing coronal field anchored in the positive polarity patch and shearing plasma motions of same positive polarity features appear to enhance the flux concentration forming the pore and making it more prominent (Figure~\ref{f4} and \ref{f6}). 
In this context, we observe that horizontal flows revealed both in the chromosphere and the photosphere seem to be reminiscent of diffusion motions detected in supergranules, as those found by \cite{Roudier2014,Roudier2016} during magnetic network formation and \cite{Chenxi2024} during sunspot decay phase. This coupling has been also studied in emerging flux regions by e.g. \cite{Georgoulis2002,Schmieder2004,Zhao2024}.

In addition, we show that the various phenomena observed at different atmospheric heights in the pore formation region are spatially and temporally linked to each other (Figures~\ref{f7} and ~\ref{f8}). The brightness enhancements are first seen in the IBIS Ca~II line core observations, at the time the flux started to increase in the photosphere at 13:00~UT. Chromospheric surges are seen in the Ca~II line wing data between 14:00 and 15:00~UT and EUV surges are identified in the SDO/AIA at 16:45 UT with some 90 min delay (Figure~\ref{f9}). SDO/AIA images also show  brightening events with plasma flows localized along the field lines of coronal loop systems. Brightness enhancement decreases with increasing ion temperature and increasing formation height (panels c and d of Figure~\ref{f3}). All these findings point to the emergence process as the likely trigger of reconnection events, due to the interaction between the pre-existing magnetic arcades and the EFR domain. Plasma motions are clearly detected in the region around the EFR and the brightness enhancements reported above just lie where some of these motions are converging. 

In this perspective, it is worth recalling that \cite{Torok2009} modeled the emergence of a twisted flux tube into a potential field arcade with a two-step process that resembles the one observed in the studied data. In particular, those authors modeled the boundary-driven "kinematic" emergence of a compact, intense, and uniformly twisted flux tube into a potential field arcade that overlies a weakly twisted coronal flux rope. The expansion of the emerging flux in the corona gives rise to the formation of a null-point at the interface between the emerging and pre-existing fields. In the two-step reconnection process, the first reconnection involves emerging fields and a set of large-scale arcade field lines. The second reconnection occurs between these newly formed loops and remote arcade fields, and yields the formation of a second loop system on the opposite side of the emerging flux. The two loop systems collectively display an anemone pattern (e.g., \cite{Reetika:24}) that is located below the fan surface. The flux that surrounds the inner spine field line of the null-point retains a fraction of the emerged twist, while the remaining twist is evacuated along the reconnected arcades. The nature and timing of the features which occur in the simulation do qualitatively reproduce those observed in X-rays in the particular event studied by \cite{Torok2009}, by suggesting the two-step reconnection process to be a consistent and generic model for the formation of anemone regions in the solar corona. Also recent numerical simulations \cite{nobrega2022} show the formation of X-ray coronal bright points above the EFR sites, due to interaction with the ambient field.

Benefiting from coordinated observations by IRIS and \textit{Hinode} satellites, \cite{toriumi2017} investigated local heating events occurring in the lower atmosphere during the earliest phase of flux emergence of a $\sim 30\arcsec \times 30\arcsec$ EFR in AR NOAA~12401, containing two $\sim 5\arcsec$ pores. They found that some Ca~II~H bright points in the EFR center, located in correspondence of mixed polarities in the photosphere, have \textit{IRIS} UV spectra that exhibit flare-like light curves and signatures of high-speed ($\pm 150 \,\mathrm{km \,s}^{-1}$) bi-directional jets. Similarly, \cite{centeno2017} analysed \textsc{Sunrise} observations that detail the emergence of a tiny ($\sim 5\arcsec$) EFR in the photosphere and its chromospheric response, which consists of brightening along the fibrils that form the AFS developing between the footpoints of the EFR. They also observed UV enhancements at 160~nm and 170~nm in the location of the EFR, possibly linked to the occurrence of Ellerman bombs in the emergence site.

However, it is worth emphasizing that with respect to the above studies we consider processes occurring on larger spatial scales and longer temporal domains. 
%
Likewise, taking advantage of high-resolution ground-based observations acquired with the GREGOR telescope, \cite{Yadav19} studied an EFR forming pores. They found in the chromosphere supersonic downflows of about 40 km/s appearing near the loop footpoints connecting two pores of opposite polarity, whereas strong upflows of 22 km/s appeared near the apex of the loops. Using non force-free field extrapolation they reconstructed the magnetic field topology of the EFR highlighting the presence of small-scale loops beneath the large loops, which could trigger various heating events due to magnetic reconnection. GREGOR observations were also used in the study of the evolution of an AFS performed by \cite{Manrique2018}. During the formation of this AFS, the authors found at the apex of the loop upflows in the range of $5 - 6$ km/s. After 4 minutes, supersonic downflows ($20 - 40$ km/s) set in while the AFS expanded by about 7\arcsec. As already reported in \cite{Lagg2007}, they observed persistent chromospheric supersonic downflows reaching values up to 40 km/s for the growing pore at the edge of the AFS. 

Finally, our work can be also compared to the findings of \cite{Verma20}. Analyzing VTT observations, they observed an AR that started as a bipolar region with continuous flux emergence when a new flux system emerged in the leading part, resulting in two homologous surges. While flux cancellation at the base of the surges provided the energy for ejecting the cool plasma, strong proper motions of the leading pores changed the magnetic field topology, making the region susceptible to surging. Despite the surge activity in the leading part, an AFS in the trailing part of the old flux remained stable. Thus, stable and violently expelled mass-loaded ascending magnetic structures can coexist in close proximity. Moreover, they found significantly broadened H${\alpha}$ profiles, implying heating at the base of the surges, which is also supported by bright kernels in UV and EUV images uncovered by swaying motions of dark fibrils at the base of the surges. In light of this, \cite{Verma20} also invoked the interaction of newly emerging flux with pre-existing flux concentrations of a young, diffuse active region as the cause that provided suitable conditions for the observed surges.

\section{Conclusions}


Pores observed in the solar photosphere constitute the first stage of the evolution of sunspots, which form ARs that can host eruptive events affecting the whole heliosphere.

We studied the evolution of an EFR in AR NOAA~11462, by focusing on the processes that led to the formation of a nearly symmetric, funnel-shaped pore. A first investigation, presented in Paper~I, addressed the pore formation in the photosphere, by tracking its development from the early stages with the analysis of the plasma motions and magnetic field configuration. Here, to provide a detailed description of the connection between the EFR and overlying atmospheric layers and of the relevance of this connection to the pore formation, we analysed simultaneous ground-based high-resolution IBIS observations of the EFR in the lower atmosphere, concentrating our attention to the chromosphere, and complementing the analysis of these observations with space-borne data acquired by the SDO satellite.

It is worth highlighting that the evolution of the events analysed in our study occurs at small spatial scales and in short time, as for similar events in the solar atmosphere, and, hence, this makes the measurements available for our analysis uncommon. This is why we described our observations extensively. Besides, the studied EFR could be investigated by exploiting several co-temporal dataset sampling the solar atmosphere from the low photosphere to the upper corona with several upcoming new facilities.

The high-cadence, comprehensive view of the evolving region manifests the magnetic and energy coupling of the solar atmosphere through atmospheric structuring, magnetic fields and connection of the physical processes, as well as the role of the magnetic fields in the heating and dynamics of the solar atmosphere. Most of the reported events are evidently causally related, and the observed coupling originates by the interplay between magnetic field and plasma motions in the EFR and in the pre-existing  magnetic environment. 

As a remarkable example, this suggests surges resulted from interaction between minority polarity flux and adjacent unipolar flux concentrations from the pre-existing magnetic environment and the coalesced flux of the EFR, due to the convergence of opposite magnetic flux patches. Moreover, the chromospheric surges are seen almost co-spatial with the field lines produced by the convergence of arcades overlying the EFR and pre-existing loops. In this perspective, the dynamic interplay between emerging fields and pre-existing field systems plays an important role for the energy coupling of the atmosphere. In fact, the rotational motions in counter-clockwise direction progressively trigger the reconnection causing the surges, whose location exhibits similar progressive displacements, as found in three-dimensional models of slipping reconnection \cite{Aulanier:06,Pariat:09}. 

Brightening events with plasma flows are localized along field lines of AFS and coronal loop systems. Intensity enhancement decreases with increasing ion temperature and increasing formation height. All these findings point to the emergence process as the likely trigger of reconnection events, due to the interaction of the pre-existing arcades and the EFR field lines, according to a bottom-up mechanism.

The use of 
spectropolarimetric imagers, like the present \textsc{Sunrise}~III TUMAG \cite{Tumag2022}, SST/CRISP \cite{CRISP2008}, DKIST/VTF \cite{VTF2016}, and the new generation instruments like TIS/FBI under development for the European Solar Telescope (EST, \cite{EST2022}), are expected to progress our understanding of the interaction of new magnetic flux with preexisting fields during AR assembly. In this frame, it worth highlighting that the IBIS instrument that provided the data analyzed in this study will be soon replaced with an upgraded version installed at a telescope of the Teide Observatory in Canary Islands \cite{IBIS2.0:20,IBIS2.0:22,IBIS2.0:24}. This instrument will get data 
in synergy with the future observations of the SOLAR-C \cite{Solarc:19} and MUSE \cite{MUSE:20,MUSE:22} missions. Using their complementary capabilities, these missions will provide both images and spectroscopic information for the investigation of the response of the transition region and coronal layers to the energy release phenomena occurring as a consequence of flux emergence.

\vspace{6pt}

\authorcontributions{Conceptualization, I.E. and S.L.G.; methodology, M.M. and I.E.; software, I.E. and M.M.; validation, S.L.G., I.E. and M.M.; formal analysis, I.E. and M.M.; investigation, I.E.; resources, I.E.; data curation, F.G. and I.E.; writing---original draft preparation, I.E.; writing---review and editing, S.L.G and M.M.; visualization, I.E. and M.M.; supervision, I.E.; project administration, I.E.; funding acquisition, I.E. All authors have read and agreed to the published version of the manuscript.}

\funding{
This study has received funding from the European Union's Horizon 2020 research and innovation program under the grant agreements No.~739500 (project PRE-EST) and No.~824135 (project SOLARNET). This work was also supported by the INAF Istituto Nazionale di Astrofisica Science Directorate (PRIN-INAF-2014) and Italian MIUR (PRIN-2012). The IBIS2.0 project is supported by he INAF Science Directorate. S.L.G, and P.R. acknowledge support from ASI/INAF agreement n. 2022-29-HH.0 Missione MUSE.
M.M. has been supported by the ASI-INAF agreement n. 2022-14-HH.0. The authors acknowledge support from the 
ASI/INAF agreement n. 2021-12-HH.0 “Missione Solar-C EUVST–Supporto scientifico di Fase B/C/D; Addendum N. 2021-12-HH.1-2024”. 

}

\dataavailability{The data analysed in this study are available at the IBIS$-$A archive, see Sect. 2 for details. 



}

\acknowledgments{The authors wish to thank Serena Criscuoli, Han Uitenbroek,  Doug Gilliam and the whole DST staff for its support during the observing campaign. 
IBIS has been designed and constructed by the INAF Osservatorio Astrofisico di Arcetri with contributions from
the Universitá di Firenze, the Universitá di Roma Tor Vergata, and upgraded with
further contributions from National Solar Observatory (NSO) and Queens University Belfast. IBIS was ran with support of the NSO, which is operated
by the Association of Universities for Research in Astronomy, Inc., under cooperative agreement with the National Science Foundation. The SDO/HMI data used in this paper are courtesy of
NASA/SDO and the HMI science team. Use of NASA’s Astrophysical Data System is gratefully acknowledged.

}

\conflictsofinterest{The authors declare no conflicts of interest.}

\begin{adjustwidth}{-\extralength}{0cm}

\reftitle{References}


\bibliography{biblio_3110.bib}

\PublishersNote{}
\end{adjustwidth}
\end{document}